# Neptune long-lived atmospheric features in 2013-2015 from small (28-cm) to large (10-m) telescopes

R. Hueso[1*], I. de Pater[2-3], A. Simon[4], A. Sánchez-Lavega[1], M. Delcroix[5], M. H. Wong[2], J. W. Tollefson[2], C. Baranec[6], K. de Kleer[7], S. H. Luszcz-Cook[8-9], G. S. Orton[10], H. B. Hammel[11], J. M. Gómez-Forrellad[12], I. Ordoñez-Etxeberria[1], L. Sromovsky[13], P. Fry[13], F. Colas[14], J. F. Rojas[1], S. Pérez-Hoyos[1], P. Gorczynski[15], J. Guarro[15], W. Kivits[15†], P. Miles[15], D. Millika[15], P. Nicholas[15], J. Sussenbach[15], A. Wesley[15], K. Sayanagi[16], S. M. Ammons[17], E. L. Gates[18], D. Gavel[18], E. Victor Garcia[17], N. M. Law[19], I. Mendikoa[1], R. Riddle[20]

[1] Universidad del País Vasco UPV/EHU, Bilbao, Spain.

[2] University of California, Berkeley, CA, USA.

[3] Delft University of Technology, Delft, The Netherlands.

[4] NASA Goddard Space Flight Center, Greenbelt, MD, USA.

[5] Commission des Observations planétaires, Sociéte Astronomique de France, Paris, France.

[6] Institute for Astronomy, University of Hawaiʻi at Mānoa, Hilo, HI. USA.

[7] SRON, Netherlands Institute for Space Research, Utrecht, The Netherlands.

[8] American Museum of National History, New York, NY, USA.

[9] Columbia University, New York, NY 10027, USA

[10] Jet Propulsion Laboratory, CA, USA.

[11] Association of Universities for Research in Astronomy, Washington DC., USA.

[12] Fundació Observatori Esteve Duran, Seva, Spain.

[13] University of Wisconsin, Space Science and Engineering Center, Madison, Wisconsin, USA.

[14] IMCCE, Observatoire de Paris, Paris, France.

[15] International Outer Planet Watch, Planetary Virtual Observatory Laboratory, Bilbao, Spain.

[16] Hampton University, Hampton, VA, USA.

[17] Lawrence Livermore National Laboratory, 7000 East Avenue, Livermore, CA 94550, USA.

[18] UCO/Lick Observatory, P.O. Box 85, Mount Hamilton, CA 95140, USA.

[19] University of North Carolina, Chapel Hill, NC, USA.

[20] Division of Physics, Mathematics and Astronomy, California Institute of Technology, Pasadena, CA, USA.

[†] Deceased.

(*) Corresponding author: Ricardo Hueso.  e-mail: ricardo.hueso@ehu.eus

**Abstract:** Since 2013, observations of Neptune with small telescopes (28-50 cm) have resulted in several detections of long-lived bright atmospheric features that have also been observed by large telescopes such as Keck II or Hubble. The combination of both types of images allows the study of the long-term evolution of major cloud systems in the planet. In 2013 and 2014 two bright features were present on the planet at southern mid-latitudes. These may have merged in late 2014, possibly leading to the formation of a single bright feature observed during 2015 at the same latitude. This cloud system was first observed in January 2015 and nearly continuously from July to December 2015 in observations with telescopes in the 2-10-m class and in images from amateur astronomers. These images show the bright spot as a compact feature at -40.1 ± 1.6° planetographic latitude well resolved from a nearby bright zonal band that extended from -42° to -20°.The size of this system depends on wavelength and varies from a longitudinal extension of 8,000±900 km and latitudinal extension of 6,500±900 km in Keck II images in H and Ks bands to 5,100±1400 km in longitude and 4,500±1400 km in latitude in HST images in 657 nm. Over July to September 2015 the structure drifted westward in longitude at a rate of 24.48±0.03°/day or -94 ± 3 m/s. This is about 30 m/s slower than the zonal winds measured at the time of the Voyager 2 flyby. Tracking its motion from July to November 2015 suggests a longitudinal oscillation of 16° in amplitude with a 90-day period, typical of dark spots on Neptune and similar to the Great Red Spot oscillation in Jupiter. The limited time covered by high-resolution observations only covers one full oscillation and other interpretations of the changing motions could be possible. HST images in September 2015 show the presence of a dark spot at short wavelengths located in the southern flank (planetographic latitude -47.0°) of the bright compact cloud observed throughout 2015. The drift rate of the bright cloud and dark spot translates to a zonal speed of -87.0 ± 2.0 m/s, which matches the Voyager 2 zonal speeds at the latitude of the dark spot. Identification of a few other features in 2015 enabled the extraction of some limited wind information over this period. This work demonstrates the need of frequently monitoring Neptune to understand its atmospheric dynamics and shows excellent opportunities for professional and amateur collaborations.

1. Introduction

Early studies of the planet Neptune showed that, in spite of its large distance to the Sun, and unlike Uranus, its atmosphere is very dynamic with several sources of variability (Belton et al., 1981; Hammel, 1989; Ingersoll et al. 1995 and references therein). Historically, the small angular size of Neptune (maximum diameter of 2.3'') resulted in a lack of spatially resolved observations of the planet until the arrival of the Voyager 2 in 1989 (Smith et al., 1989). The launch of the Hubble Space Telescope (HST) and the development of high performance Adaptive Optics (AO) on large ground-based telescopes allowed monitoring the atmospheric activity of the planet at high resolution. Neptune shows rapidly varying cloud activity, zonal bands that change over the years, long-lived dark ovals and sporadic clouds around them (e.g. Limaye and Sromovsky, 1991; Baines et al. 1995; Ingersoll et al., 1995; Karkoschka, 2011; Sromovsky et al., 2001b, 2002; Fry and Sromovsky, 2004; Martin et al., 2012.; Fitzpatrick et al., 2014). Recently, spatially resolved observations of the planet have also become possible at thermal infrared (Orton et al., 2007; Fletcher et al.. 2014), millimeter (Luszcz-Cook et al., 2013) and radio wavelengths (de Pater et al., 2014) opening the possibility to study the thermal structure of the stratosphere and the structure of the troposphere below the visible clouds (de Pater et al., 2014).

Neptune's global circulation is dominated by a broad retrograde westward equatorial jet with a peak velocity of 350 m/s that diminishes at higher latitudes until it gives way to prograde winds from mid-latitudes to the South Pole (and presumably also in the North polar region not yet observed) in a narrower eastward jet with a peak velocity of +300 m/s at -74°S (Sánchez-Lavega et al., 2017). These winds were first measured in images at visible wavelengths from the Voyager 2 spacecraft in its flyby of the planet in 1989 (Stone and Miner, 1991; Limaye and Sromovsky, 1991; Sromovsky et al., 1993). Later wind measurements were obtained from images acquired by Adaptive Optics instruments on the Keck II telescope operating in the near infrared at 1-2.3 µm (Fry and Sromovsky, 2004; Martin et al., 2012; Fitzpatrick et al., 2014,

Tollefson et al., 2016) and from HST images in the visible (e.g. Hammel and Lockwood, 1997; Sromovsky et al., 2001b, 2002). These observations are sensitive to clouds and hazes from 0.1 to 0.6 bar (Fitzpatrick et al., 2014). Zonal wind profiles from those measurements are generally consistent with the one derived from Voyager 2. However there is a large dispersion of velocities in analysis of features tracked over short time periods when compared to the Voyager results (Limaye and Sromovsky, 1991; Martin et al., 2012). Part of this variability might be caused by vertical wind shear (Martin et al, 2012; Fitzpatrick et al., 2014), specially close to the Equator where vertical wind shear can be on the order of 30 m/s per scale height from Voyager IRIS data (Conrath et al., 1989) and similarly from IR data in 2003 (Fletcher et al., 2014). Most of this variability seems linked to the different apparent motions of bright and large features observed over long time-scales compared with smaller and fainter clouds observed only for a few hours and in many cases affected by their interaction with large features nearby. Therefore, sources of variability in zonal wind measurements include intrinsic variability of the small clouds, vertical wind shear, and the short time differences from consecutive images used for some measurements that introduce uncertainties that add to the real variability.

Studies of Voyager-2 images in the visible (Baines et al., 1995) and recent observations in the near infrared and at radio wavelengths (de Pater et al., 2014) conclude that the overall cloud structure of the planet consists of different vertical layers that vary with latitude and time (Irwin et al., 2016). The main cloud deck level (made of methane ice crystals), observed only at wavelengths not sensitive to methane absorption is estimated to lie at around the 2–3 bar level (Irwin et al., 2011). At southern tropical to mid-latitudes a belt of hazes, visible in methane absorption bands, is located at P ~300-600 mbar and is overcast intermittently by a stratospheric haze possibly made of condensed hydrocarbons at 20-30 mbar that could arise from changing temperatures or from materials brought up from the troposphere. The latitudinal position, overall activity and latitudinal extension of this bright belt changes from year to year. For instance, in 2013 it extended roughly from -45° to -27° with diffuse latitudinal limits and it acquired a more compact structure in 2015 with a latitudinal size from -42° to -21°. At northern mid-latitudes the main cloud seems quite similar while the stratospheric clouds seem to be located a bit higher, near 10 mbar. Based on maps of the thermal emission and hydrocarbons abundances in Neptune's stratosphere obtained from Voyager observations in the mid infrared, Conrath et al. (1991) and Bézard et al. (1991) proposed a global circulation of the atmosphere with rising cold air at mid latitudes and overall descent at the Equator and the polar latitudes. This global circulation has been further explored to explain also the cloud structure in the planet by de Pater et al. (2014). This overall structure matches Neptune's distribution of ortho/para hydrogen and thermal structure at the time of the Voyager-2 encounter (heliocentric longitude $L_S$ = 236°, Conrath et al., 1989). It also matches the visual aspect at near infrared wavelengths (1.2-2.3 μm) for the last few years, which is characterized by bright belts of clouds at northern and southern mid-latitudes as well as occasionally bright south polar features. This visual aspect of the planet corresponds to early autumn in the south hemisphere (southern summer solstice was in 2005 and heliocentric longitudes from 2013 to 2015 were 287° to 293°). An analysis of vertical wind shear in the equatorial region, however, is consistent with upwelling at P>1 bar, suggesting a more complex circulation pattern, such as a stacked-cell circulation with reversed flow above and below 1 bar (Tollefson et al. 2016).

A challenge to our understanding of the atmosphere is the sparse temporal sampling of high-resolution images of the planet. Much better temporal sampling has been achieved within the last few years by amateur astronomers using small telescopes of 50 cm or smaller to monitor some of Neptune's atmospheric features (Delcroix et al., 2014b). This revolution in observations of the icy giants has been enabled by the use of new fast CCD cameras with improved sensitivity in the near infrared (650 – 1000 nm) and low-cost long-pass imaging filters, typically starting at 610 – 700 nm and extending until the detector cutoff close to 1 μm (Mousis et al., 2014).

Here we present observations of recent cloud activity from July 2013 to December 2015 (sub-solar latitudes moving from -26.9° in July 2013 to -25.9° in December 2015). We analyze data provided by telescopes with diameters from 28 cm (amateur size) to 10-m (Keck II telescope), including data from HST, performing a long-term tracking of the brightest atmospheric features. We present a description of bright features in 2013, 2014 and 2015. We present a description of our observations in section 2. Image navigation and measurement techniques are presented in section 3. Analysis of the images is presented in section 4. Section 5 presents drift rates of the main cloud features in terms of zonal winds comparing with previous studies. Finally, we present a summary of our findings and conclusions in section 6.

**2. Observations**

2.1. 2013 images

Images of Neptune in the visible range obtained at the 1.06-m telescope at Pic-du-Midi (France) in July 2013 showed a bright cloud feature at southern mid-latitudes. This telescope is frequently used by French amateur astronomers using low-cost commercial imaging cameras (Delcroix et al., 2014a). Due to recent advancements in affordable cameras with high Quantum Efficiency in the short infrared, the bright feature on Neptune was later confirmed by several observers using telescopes of 28-38 cm in August and September 2013. This was the first time that amateur astronomers could repeatedly observe the same cloud feature on the planet. About 13 amateur observations showed features on the planet. Six observations showed a bright feature at approximately the same latitude (-45° planetographic) while other candidate spots at other planetographic latitudes from +2° to -73° could not be confirmed on a sequence of images (Delcroix et al. 2014b). Most of the amateur images used in this study are publicly available in the Planetary Virtual Observatory and Laboratory (PVOL) database (Hueso et al., 2010; 2017) database available on http://pvol2.ehu.eus) and form part of the International Outer Planets Watch (IOPW)-Atmospheres collaboration.

Images obtained in October 2013 at Calar Alto Observatory in Spain using the 2.2-m telescope and the AstraLux camera (Hormuth et al., 2008) also showed two features on the planet at the same latitude although with a low contrast. Fig. 1 shows the 2013 Pic du Midi image, representative examples of amateur observations, and one of the observations obtained at Calar Alto. All images included Triton as a reference to orient the images and measure the position of atmospheric features.

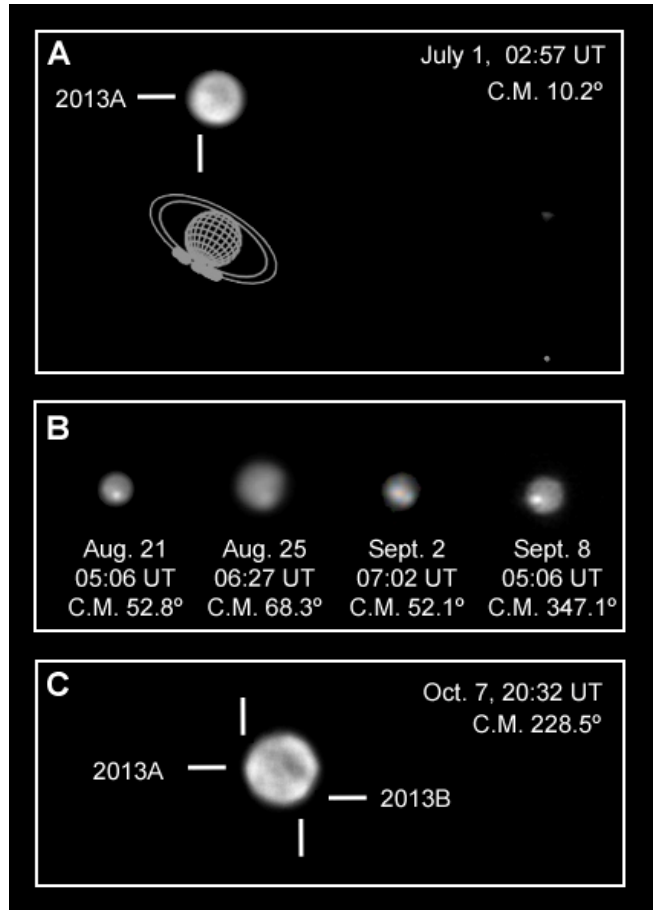

**Figure 1: Large bright features in Neptune in 2013**. (A) Pic du Midi first observation of a bright feature in Neptune. (B) Amateur observations on different dates. (C) Calar Alto observation in October 7. This last image shows two highlighted features in both limbs and a north bright limb. All the images are oriented in the same sense as shown in the upper panel and using Triton's position as a reference. North is up and West is to the left with the planet tilted as it appears on the sky. Observer names, filters and details are given in Table 1. Features discussed in sections 4 and 5 are labeled in the figure.

Keck II observed the planet the $3^{rd}$ and $31^{st}$ of July with the NIRC2 (1-5.0 μm) camera. Tollefson et al. (2016) present an in depth analysis of this dataset. ESO's Very Large Telescope (VLT) also observed Neptune on 9 to 12 October 2013 using the SINFONI instrument (Eisenhauer et al., 2003) that operates in the spectral range of 1.1 to 2.5 µm and the observations were analyzed by Irwin et al. (2016). Representative examples of these observations appear in Fig. 2, which are difficult to compare with amateur observations because of different spatial resolutions and very different wavelengths: the amateur data was obtained in the visible and up to 1 µm, whereas the Keck and VLT data covered the near infrared, which is dominated by strong methane absorption bands. VLT images showed two white bright features compatible with the two bright features observed in the Calar-Alto image and similar to the single bright feature observed in different amateur observations and Pic du Midi and Keck observations. Table 1 summarizes the characteristics of observations that detected large atmospheric features on the planet in 2013. Later analyses (section 4.1) showed that two bright features, here called 2013-A and 2013-B were observed on different dates.

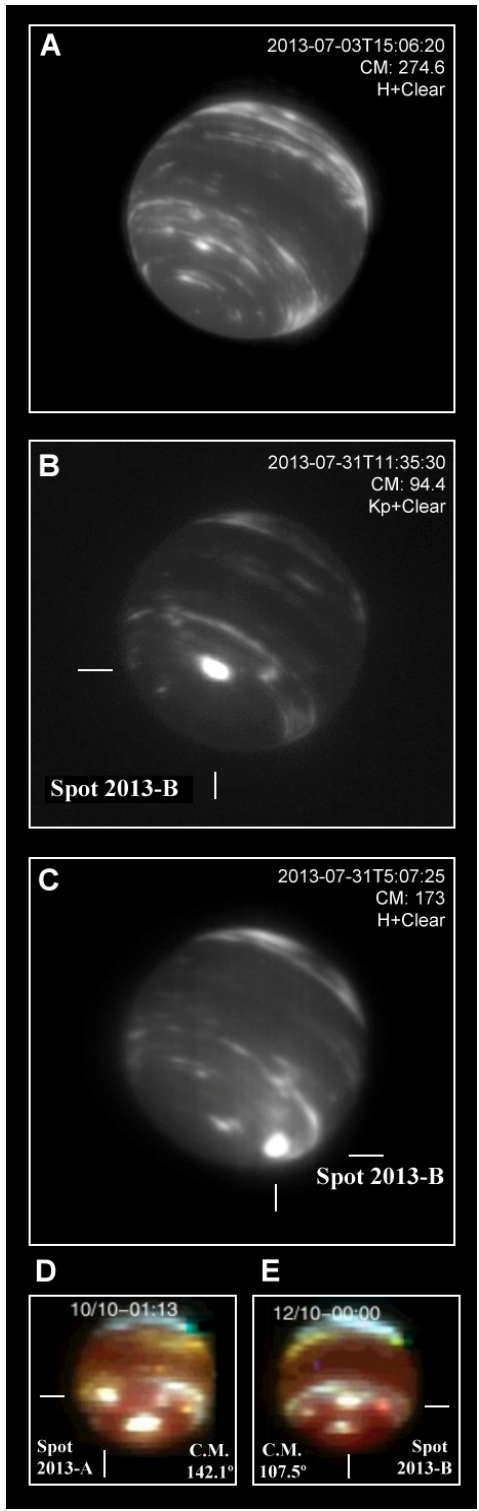

**Figure 2: Large bright features on Neptune in 2013.** Keck II (panels A, B and C) and VLT images (panels D and E) of Neptune on different dates in 2013. All images were obtained in the near infrared. Orientation on panels A to C is like in Figure 2. Panels D and E have North up and west to the left (not tilted). Colors in panels D and E roughly indicate deep clouds at 3 bar (red); intermediate clouds at 1.25 bar (green) and hazes at 0.2 bar (blue). See Irwin et al. (2016) for details on filters and altitudes on these observations. Two large features are identified in Keck and VLT images at mid-latitudes similar to the data in Fig. 1. On some dates no bright features are on the observable side of the planet. Further details are given in Table 1.

**Table 1: Neptune observations of bright features in 2013**

| Date (yyy-mm-dd) | Telescope | PI or observer | Camera / Instrument | Filter |
|---|---|---|---|---|
| 2013-07-01 | Pic du Midi (1.06 m) | F. Colas / M. Delcroix | Basler acA640-100gm | > 685 nm |
| 2013-07-31 | Keck II (10 m) | I. de Pater | NIRC2 | H, Kp |
| 2013-08-10 | Pic du Midi (1.06 m) | F. Colas / M. Delcroix | Basler acA640-100gm | > 685 nm |
| 2013-08-21 | 36 cm | P. Gorczynski | ASI 120 MM | > 685 nm |
| 2013-08-25 | 36 cm | P. Gorczynski | ASI 120 MM | > 685 nm |
| 2013-08-25 | 28 cm | J. Boudreau | ASI 120 MM | > 685 nm |
| 2013-09-02 | 36 cm | P. Maxson | ASI 120 MM | > 610 nm |
| 2013-09-08 | 38 cm | P. Jones | ASI 120 MM | > 685 nm |
| 2013-09-12 | 38 cm | P. Jones | ASI 120 MM | > 685 nm |
| 2013-10-07 | Calar Alto (2.2 m) | A. Sánchez-Lavega | AstraLux | Johnson I |
| 2013-10-10 | VLT (8.2 m) | P. Irwin (*) | SINFONI | 1.1-2.5 μm |

* VLT images acquired from 9 to 12 October and reported in Irwin et al. (2016). The single date listed here corresponds to the best visibility of the bright clouds that could be identified and related to previous observations.

2.2. 2014 images

2.2.1. Images from telescopes with diameters 0.36-1.5 m

Amateur observations of Neptune in 2014 showed the presence of at least one bright feature similar to that of 2013 at a similar latitude (-38° planetographic). The first detection of this feature was reported by P. Gorczynski in September 20, using a 36 cm telescope and was later confirmed on higher-quality observations obtained with the Pic du Midi telescope and with the instrument Robo-AO at the Palomar 1.5-m telescope. Robo-AO is an autonomous laser-adaptive optics system and science instrument designed to robotically observe at the diffraction limit in the visible (Baranec et al., 2013, 2014). Fig. 3 shows representative observations of Neptune over this period and Table 2 lists the images used for this study.

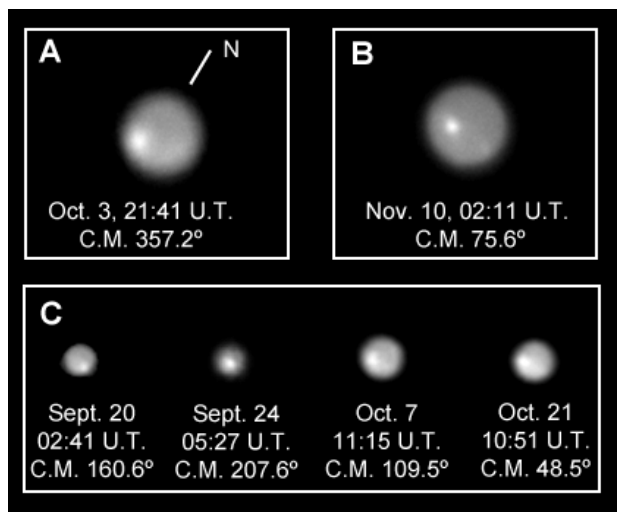

**Figure 3: Large bright features in Neptune in 2014.** (A) Pic du Midi and (B) Robo-AO observations of a bright feature in Neptune in 2014. (C) Examples of images acquired on different dates by amateur astronomers. All the images are oriented using as a reference the position of Triton (not shown). North polar direction is indicated in panel A and is the same for all the images (North is up and West is to the left with the planet tilted as it appears on the sky). Observer names, filters and details are given in Table 2.

2.2.2. Keck observations

Keck II observed the planet on August 20 (Tollefson et al., 2016). Observations in bands H (1.65 μm) and K (2.2 μm) showed two different bright features at nearby latitudes and compatible with the features that were later observed by amateur astronomers. We call these features 2014-A and 2014-B. Fig. 4 shows representative images in the H band and a partial cylindrical map of the planet.

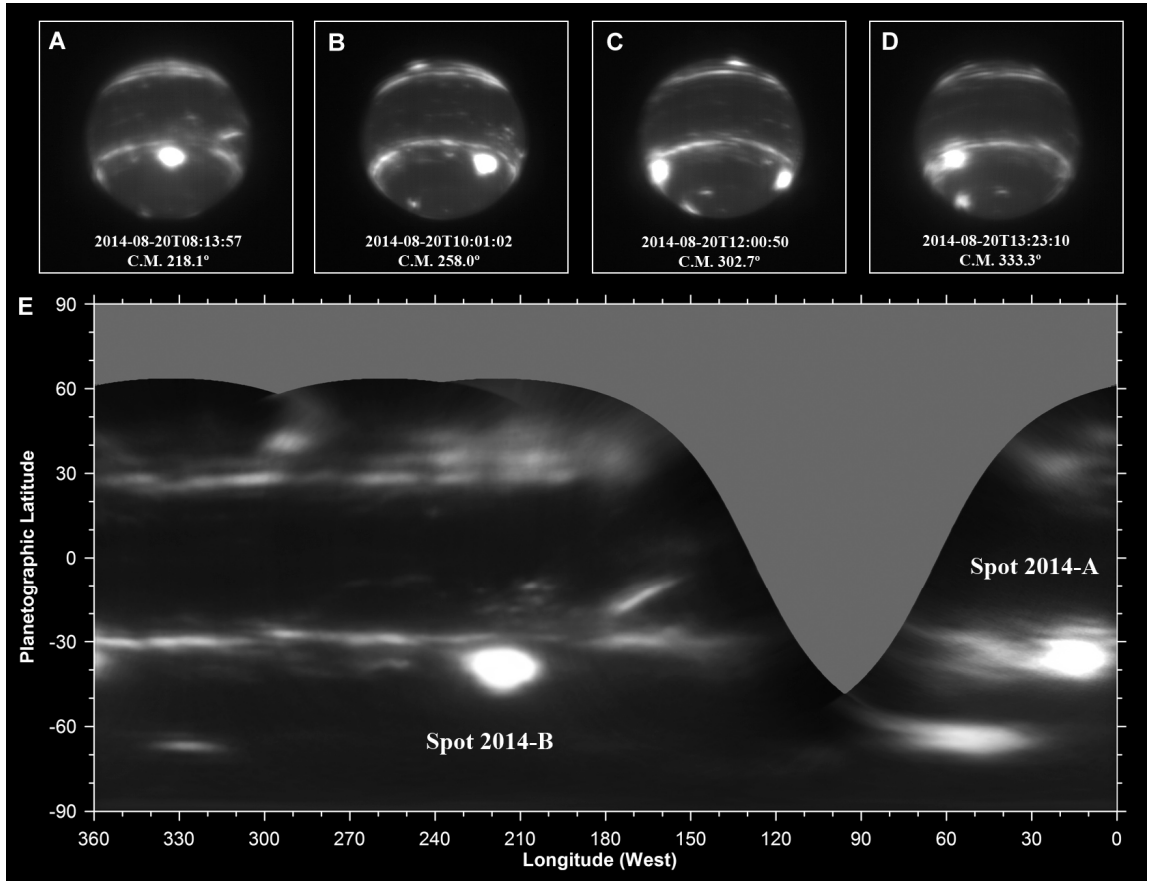

**Figure 4: Keck II NIRC2 images of Neptune on August 20, 2014 and nearly full map of the planet.** (A-D) Series of Keck II NIRC2 images of Neptune in H band (1.65 µm). North is up and West to the left (not tilted). (E) Composition of cylindrical projection of the images showing two outstanding bright features labeled.

**Table 2: Neptune observations of bright features in 2014.**

| Date (yyy-mm-dd) | Telescope and diameter | PI or observer | Camera | Filter |
|---|---|---|---|---|
| 2014-08-06 | Keck II (10 m) | I. de Pater | NIRC2 | H, Kp |
| 2014-08-20 | Keck II (10 m) | I. de Pater | NIRC2 | H, Kp |
| 2014-09-20 | 36 cm | P. Gorczynski | ASI 120 MM | > 685 nm |
| 2014-09-24 | 36 cm | P. Maxson | ASI 120 MM | > 610 nm |
| 2014-10-01 | 36 cm | P. Maxson | ASI 120 MM | > 610 nm |
| 2014-10-02 | 36 cm | P. Maxson | ASI 120 MM | > 610 nm |
| 2014-10-02 | 30 cm | N. Haigh | ASI 224 MC | > 685 nm |
| 2014-10-03 | Pic du Midi (1.06 m) | F. Colas / M. Delcroix | ASI 120 MM | > 685 nm |
| 2014-10-07 | 37 cm | A. Wesley | GS3-U3-32S4M | > 600 nm |
| 2014-10-11 | 36 cm | P. Maxson | ASI 120 MM | > 610 nm |
| 2014-10-15 | 36 cm | P. Maxson | ASI 120 MM | > 610 nm |
| 2014-10-21 | 37 cm | A. Wesley | GS3-U3-32S4M | > 600 nm |
| 2014-11-10 | Robo-AO (1.5 m) | C. Baranec | --- | g', r', i', z' |
| 2014-11-10 | 36 cm | P. Maxson | ASI 120 MM | > 610 nm |

2.3. 2015 images

2.3.1 2015 Calar Alto observations

We observed Neptune at visible and near-infrared wavelengths on 13 and 14 July, and on October 10 with the 2.2-m telescope at Calar Alto Observatory in Spain using the PlanetCam UPV/EHU dual camera instrument (Mendikoa et al., 2016). Images were acquired in several narrow and wide-band filters in the visible and the near infrared up to 1.7 μm. In July 13, 2015 a bright atmospheric feature at mid-latitudes was easily observed. We call this feature 2015-A. Figure 5 shows a selection of observations in those filters where the bright feature can be observed or even dominates the brightness of the planet. Observations at shorter wavelengths including blue filters, or in the visible, but at wavelengths without methane absorption bands, showed a bland planet without atmospheric features. Images in Fig. 5 have not been processed after stacking retaining the real contrast of the features. The contrast of the bright features varied strongly in different image filters and intensified in those with methane absorption like the 890-nm strong absorption band. The contrast peaked in the J (1.25 μm) and H (1.65 μm) bands where the bright feature alone accounted for 28 and 40% of the total light coming from the planet suggesting high cloud tops.

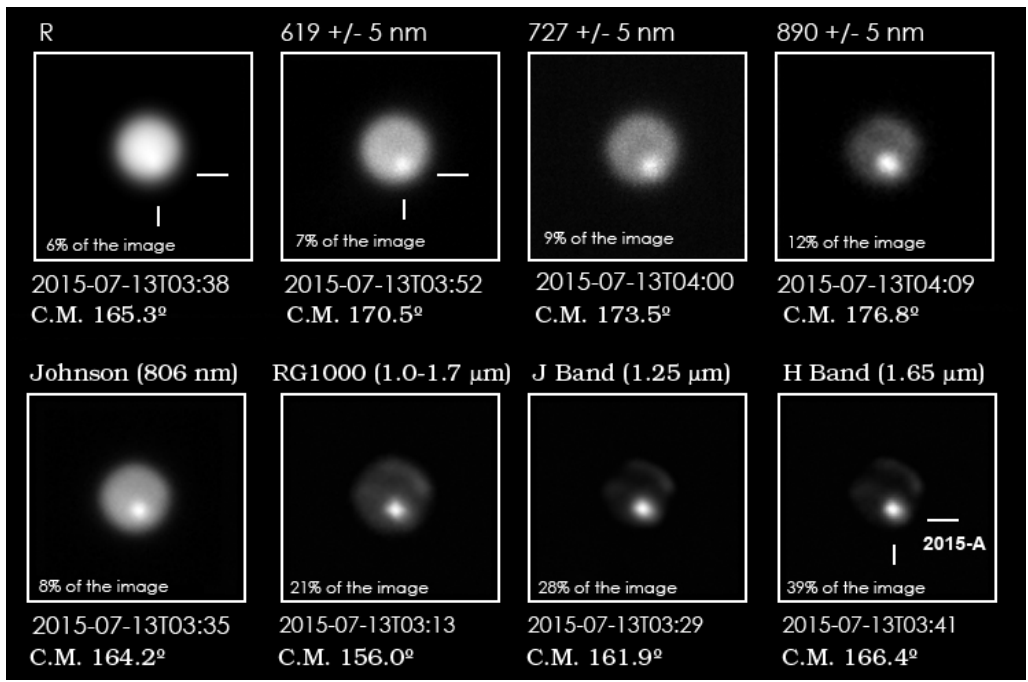

**Figure 5: Calar Alto PlanetCam observations of a bright feature in Neptune in 2015.** Filters and times are indicated in each panel. Images cover the longitudinal range from 240° (West limb) to 75° (East limb) at the latitude of the bright feature. Percentages in the insets indicate the comparative brightness of the feature with respect to the full planet. The relative brightness and contrast of this atmospheric feature increased at spectral regions progressively dominated by methane absorption. Images are oriented like in Figure 1 (North is up and West is to the left with the planet tilted).

Figure 6 shows the other side of the planet observed the following night. While several other atmospheric features are observed, their relative contrast with the rest of the planet is lower in all cases than in the bright feature observed the previous day at comparable wavelengths. Besides the regular structure of the South mid-latitudes belt, a North tropical bright feature and a South polar cloud are observable in this side of the planet.

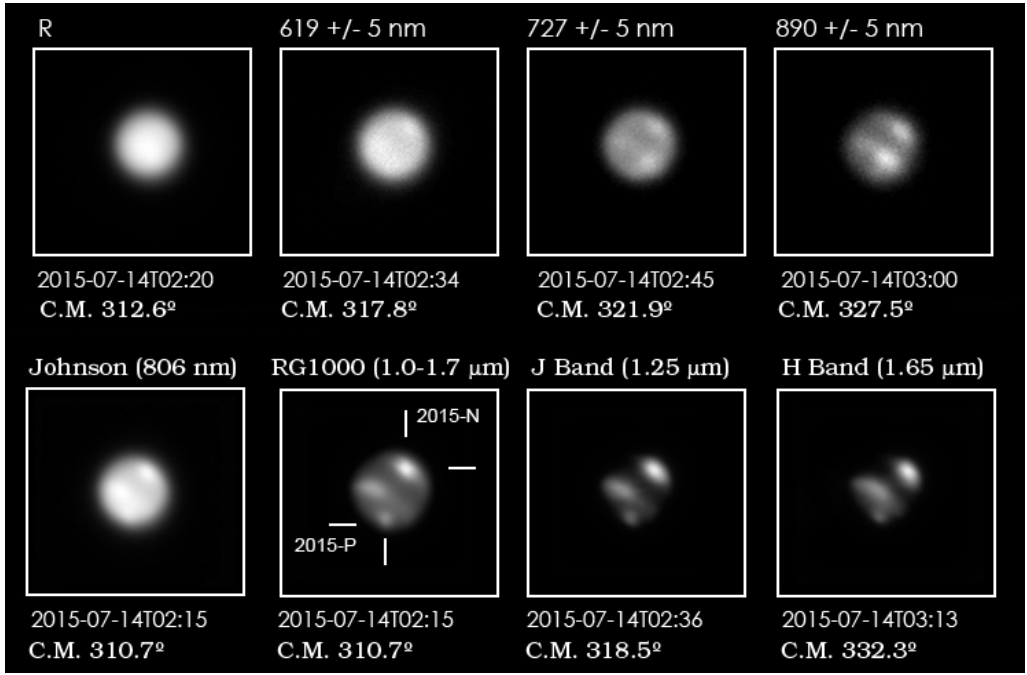

**Figure 6: Calar Alto PlanetCam observations of Neptune in 2015.** These observations were acquired one day later to those in Figure 5 showing the other side of the planet. Images cover the longitudinal range from 45° (West limb) to 230° (East limb) at the latitude of the bright feature. Atmospheric features, especially in the South mid-latitudes belt of clouds, have significantly less contrast than the bright feature observed the previous night. Other observations in nearby wavelengths and short wavelengths without methane absorption failed to show any atmospheric feature in the planet. Orientation is like in Figure 5. Features also visible in later observations in the North (2015-N) and South (2015-P) hemispheres are labeled in one of the panels.

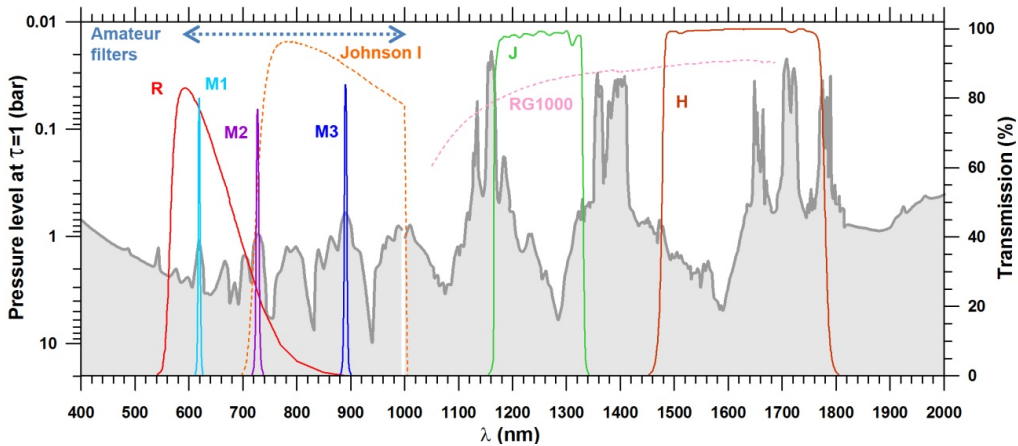

**Figure 7: Pressure levels sensed with different PlanetCam filters.** Grey solid line (left axis) shows the atmospheric penetration depth in Neptune considering the pressure level for which a two-way optical depth of unity is reached in a cloud-free model of the atmosphere. Values from 400 to 1000 nm are from Simon et al. (2016) and calculated following Sromovsky et al. (2001a). Values from 1000 to 2000 nm are from Irwin et al. (2016). The transmission of PlanetCam filters (right axis) is included for comparison (from Mendikoa et al., 2016). The cut at 1 μm of the Johnson I filter is produced by a dichroic mirror in the PlanetCam instrument. The horizontal dotted arrow indicates typical range of wavelengths covered by long-pass filters used by amateurs. Those filters sample regions where methane absorption is significant and contrast is provided by the high-altitude features.

Fig. 7 shows the pressure level for which a two-way optical depth of unity is reached in a combination of two cloud-free models of the atmosphere that include Rayleigh scattering and gas absorption. The first model is described by Simon et al. (2016) and is based on earlier calculations by Sromovsky et al. (2001a). This model provides the pressure level at which an optical opacity of 1 is reached at wavelengths from 400 nm to 1.9 μm. However the methane absorption coefficients at long wavelengths in that work were not accurate. The second model is described by Irwin et al. (2016) who used updated methane absorption coefficients and provides data from 800 nm to 2 μm. Both models are consistent in the 800-1100 nm region and a combination of both models with the data at short wavelengths (from 400 to 1000 nm) from Simon et al. (2016) and at long wavelengths (1000 nm to 2 μm) from Irwin et al. (2016) provides a reasonable estimation of the sensitivity of the observations to different vertical levels. Strong methane absorption translates into reaching optical depths of 1 at low pressures and high-altitude levels in the atmosphere. Fig. 7 also shows the transmission curves for PlanetCam filters. The contrast of the bright cloud in Fig. 5 increases at wavelengths dominated by stronger absorption bands suggesting that feature 2015-A has a cloud top at a high altitude. Parts of the spectrum covered by the J and H bands have minimum penetration depths of 0.2 and 0.03 bar respectively (Irwin et al. 2016) arguing in favor of cloud top altitudes above the 0.2 bar level for this feature similarly to typical altitudes of cloud systems in Neptune. The three narrow filters at the three methane absorption bands of 619, 727 and 890 nm have penetration depths of 1.0 to 0.6 bar in the absence of clouds and the cloud 2015-A has a very low contrast at these wavelengths.

2.3.2 Amateur observations in 2015

Amateur observations of Neptune in 2015 were far more numerous than in previous years. After the Calar Alto/PlanetCam observation showed the bright feature in Fig. 5, an alert was started in the PVOL website and a quick follow-up by some amateur astronomers engaged other observers. Fig. 8 shows examples from this campaign. A total of 20 amateurs observed a bright feature in the planet performing nearly continuous observations from July to December on 45 different. Another bright feature located in the Northern hemisphere (2015-N) and also observed in the Calar Alto images shown in Fig. 6 was observed by 8 amateurs on 9 dates. Another three amateur observations showed a south polar feature (2015-P) also present in Calar Alto images in Fig.6. The smallest amateur telescope that was able to successfully observe the bright spot 2015-A was a 28-cm refractor. Fig. 8 shows representative examples of these observations. Table 3 summarizes the dates and characteristics of the telescopes used.

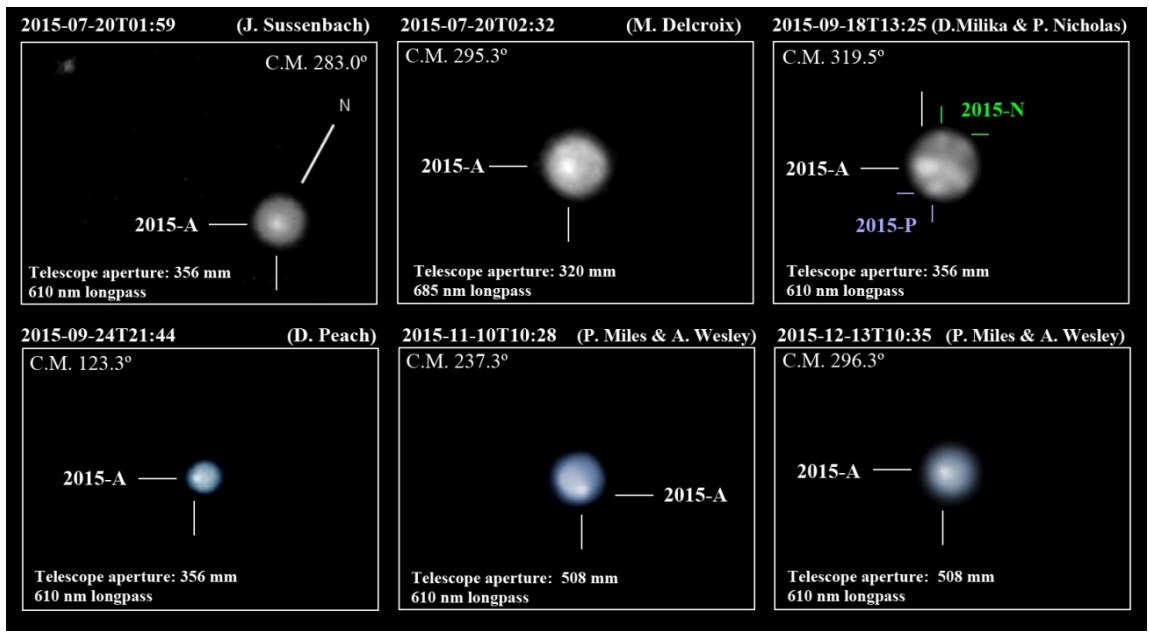

**Figure 8: Amateur observations of Neptune in 2015.** The position of Triton is shown in the first image and provides essential information to navigate these images accurately. Triton is not shown in the other images to save space. All images have the same orientation as in Fig. 1. The position of spot 2015-A is shown in all images. Some amateur observations, like the top-right image, show a wealth of other atmospheric features that can also be identified in observations in large telescopes obtained close in time.

Table 3: Neptune amateur observations of bright features in 2015

| Date (yyy-mm-dd) | Telescope diameter (cm) | Observer | Camera | Filter | Feature (†) |
|---|---|---|---|---|---|
| 2015-07-20 | 28 | A. Germano | ASI 224 MC | > 610 nm | N |
|  | 36 | J. Sussenbach | QHY5L-II | > 610 nm | A |
|  | 32 | M. Delcroix | ASI 224 MC | > 685 nm | A |
| 2015-07-21 | 32 | M. Delcroix | ASI 224 MC | > 685 nm | N |
| 2015-07-30 | 32 | M. Delcroix | ASI 224 MC | > 685 nm | A |
| 2015-07-31 | 30 | N. Haigh | ASI 224 MC | > 685 nm | N |
| 2015-08-01 | 36 | J. Sussenbach | ASI 224 MC | > 610 nm | A |
|  | 36 | W. Kivits | DMK618 | > 658 nm | A |
| 2015-08-03 | 32 | M. Delcroix | ASI 224 MC | > 685 nm | A |
| 2015-08-10 | * | A. Obukhov | * | * | A |
| 2015-08-12 | * | A. Obukhov | * | * | A |
| 2015-08-15 | 30 | A. Garbelini | ASI 224 MC | * | A |
| 2015-08-17 | 25 | S. Gonzalès | ASI 224 MC | > 610, >742 | A |
| 2015-08-20 | 25 | S. Gonzalès | ASI 224 MC | > 610, >742 | A, N |
| 2015-08-22 | * | A. Obukhov | * | * | A |
| 2015-08-24 | * | A. Obukhov | * | * | A |
| 2015-08-29 | 36 | S. Fugardi | ASI 174 MM | > 610 nm | A |
| 2015-08-30 | 28 | T. Hansen | QHY 5L-IIm | > 610 nm | A |
| 2015-08-31 | 28 | A. Germano | ASI 224 MC | > 610 nm | A |
|  | 62 | C. Pellier | PLA-Mx | > 685 nm | A |
| 2015-09-01 | 28 | T. Hansen | QHY 5L-IIm | > 610 nm | A |
|  | 36 | J. Sussenbach | ASI 224 MC | > 610 nm | A |
|  | 62 | C. Pellier | PLA-Mx | > 685 nm | A |
| 2015-09-03 | 28 | A. Germano | ASI 224 MC | > 610 nm | A |
|  | 62 | C. Pellier | PLA-Mx | > 685 nm | A |
| 2015-09-05 | 28 | T. Hansen | QHY 5L-IIm | > 610 nm | A |
| 2015-09-06 | 30 | N. Haigh | ASI 224 MC | > 685 nm | N |
| 2015-09-07 | 36 | P. Maxson | ASI 120 MM | > 610 nm | A |
| 2015-09-08 | 36 | J. Sussenbach | ASI 224 MC | > 610 nm | A |
| 2015-09-10 | 36 | W. Kivits | DMK618 | > 658 nm | A, N |
|  | 50 | S. Gonzalès | ASI 224 MC | * | A |
|  | 36 | J. Sussenbach | ASI 224 MC | > 610 nm | A |
|  | 28 | T. Hansen | QHY 5L-IIm | > 610 nm | A |
|  | 28 | A. Germano | ASI 224 MC | > 610 nm | A |
| 2015-09-12 | 36 | S. Fugardi | ASI 174 MM | > 610 nm | A |
|  | 30 | M. Miniou | Basler acA650 | > 685 nm | A |
| 2015-09-13 | 28 | A. Germano | ASI 224 MC | > 610 nm | A |
| 2015-09-17 | 30 | N. Haigh | ASI 224 MC | > 685 nm | A |
|  | 30 | Michel Miniou | Basler acA650 | > 685 nm | A |
|  | 36 | P. Maxson | ASI 120 MM | > 610 nm | A |
| 2015-09-18 | 36 | D. Milika, P. Nicholas | ASI 224 MC | > 610 nm | A, N, P |
| 2015-09-19 | 36 | D. Peach | ASI 224 MC | > 610 nm | A |
| 2015-09-20 | 36 | M. Phillips | ASI 174 MM | > 658 nm | A |
|  | 30 | Michel Miniou | Basler acA650 | > 685 nm | A |
| 2015-09-22 | 28 | A. Germano | ASI 224 MC | > 610 nm | A |
| 2015-09-24 | 36 | P. Maxson | ASI 120 MM | > 610 nm | A |
|  | 36 | D. Peach | ASI 224 MC | > 610 nm | A |
| 2015-09-25 | 36 | P. Maxson | ASI 120 MM | > 610 nm | A |
| 2015-10-01 | 36 | P. Maxson | ASI 120 MM | > 610 nm | A |
| 2015-10-04 | 36 | W. Kivits | DMK618 | > 658 nm | A |
| 2015-10-06 | 36 | W. Kivits | DMK618 | > 658 nm | A |
| 2015-10-08 | 30 | N. Haigh | ASI 224 MC | > 685 nm | N |
| 2015-10-11 | 36 | W. Kivits | DMK618 | > 658 nm | A, N |
| 2015-10-11 | 36 | J. Sussenbach | ASI 224 MC | > 610 nm | N |
| 2015-10-14 | 36 | D. Milika, P. Nicholas | ASI 224MC | > 610 nm | A, P |
| 2015-10-20 | 36 | W. Kivits | DMK618 | > 658 nm | A |
| 2015-11-01 | 36 | W. Kivits | DMK618 | > 658 nm | A |
| 2015-11-09 | 36 | P. Maxson | ASI 120 MM | > 610 nm | A |
| 2015-11-10 | 36 | A. Wesley | GS3-U3-32S4M | > 600 nm | A |
| 2015-11-15 | 36 | D. Milika, P. Nicholas | ASI 224MC | > 610 nm | A |
| 2015-11-17 | 51 | P. Miles, A. Wesley | GS3-U3-32S4M | >610 nm | A |

| 2015-11-22 | 51 | P. Miles, A. Wesley | GS3-U3-32S4M | >610 nm | A |
| 2015-11-24 | 51 | P. Miles, A. Wesley | GS3-U3-32S4M | >610 nm | A |
| 2015-11-25 | 36 | W. Kivits | DMK618 | > 658 nm | A |
| 2015-12-07 | 51 | P. Miles, A. Wesley | GS3-U3-32S4M | >610 nm | A |
| 2015-12-13 | 51 | P. Miles, A. Wesley | GS3-U3-32S4M | >610 nm | A |
| 2015-12-31 | 36 | W. Kivits | DMK618 | > 658 nm | A |

(†) A stands for feature 2015-A; N stands for feature 2015-N; P stands for feature 2015-P.
(*) Unknown.

2.3.3. Observations from the Hale and Shane telescopes at Palomar and Lick observatories

We acquired several observations with the Hale 5.1-m telescope at Palomar Observatory using the PALM-3000 Adaptive Optics system (Dekany et al. 2013) and the Project 1640 (P1640) instrument (Hinkley et al. 2011) with the calibration wave-front sensor off. We also acquired observations with the Shane 3-m telescope at Lick observatory using its ShaneAO/ShARCS Adaptive Optics instrument (Gavel et al., 2016) in H (1.65 μm) and Ks (2.17 μm) bands. Fig. 9 shows images that captured the spot 2015-A. A map of the morphology of the bright feature from Palomar images shows a distinct shape that will be compared to other high resolution observations of the same feature in section 4.

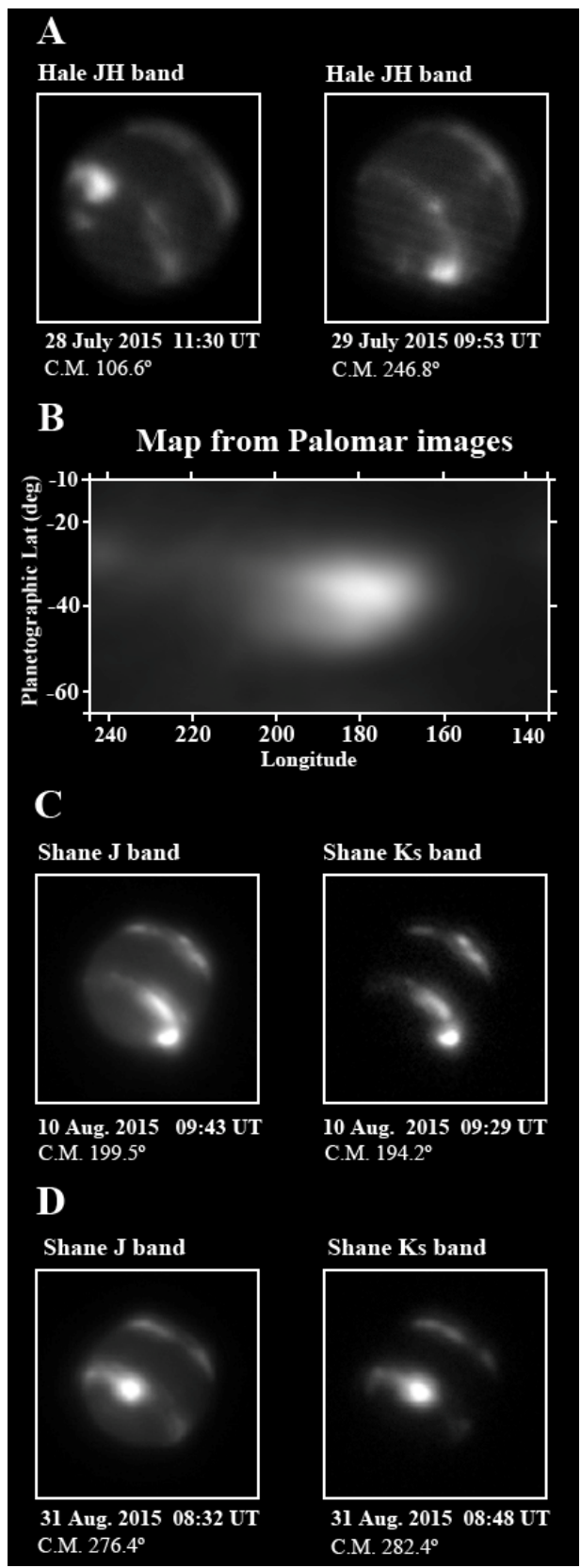

**Figure 9: Observations from Palomar and Lick observatories running AO instruments.** (A) Palomar Hale telescope data in JH bands. (B) Map of the bright feature obtained from both Palomar Hale images. (C and D) Observations in bands H and Ks with the Shane telescope at Lick observatory. Images are oriented like in Fig. 1.

2.3.4. Keck II NIRC2 observations

Images were acquired with the NIRC2 AO camera in the Keck II telescope in 25 July, 5 August, 29 August and 30 August 2015. Fig. 10 shows examples of the images together with a nearly full map of the planet. Several other atmospheric structures are seen with different brightness levels, including north tropical and south polar cloud features.

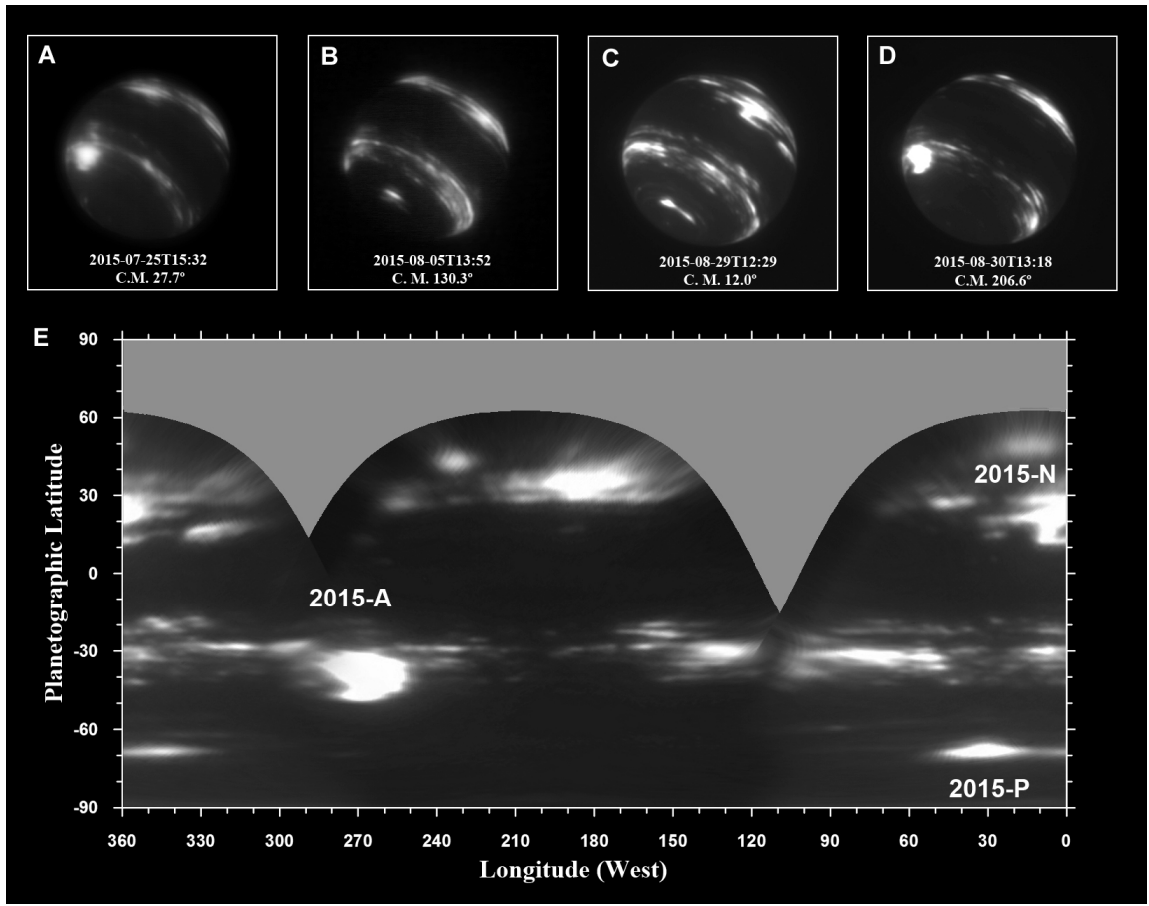

**Figure 10:** (A-D) Series of Keck II observations of Neptune in band H over July and August 2015 showing the bright spot on two different dates and other systems of clouds. Images oriented like in Fig. 1. (E) Cylindrical map of the planet from images acquired in 29 and 30 August. Note the shape of the bright spot 2015-A similar to the shape observed in Palomar images.

Additionally, Keck II observed a single large-size mid-latitudes bright spot in January 2015 (see Fig. 3 in Simon et al., 2016), implying that the bright 2015-A feature was present in Neptune's atmosphere from January to December 2015.

2.3.5. HST observations

HST observed Neptune during two periods in 2015. Images were acquired on 1 and 2 September 2015 (PI: de Pater) and on 18 and 19 September 2015 as part of the Outer Planets Atmosphere Legacy Program (OPAL: PI: A. Simon). Simon et al. (2016) report these HST/OPAL Neptune observations with a focus on the extraction of full integrated light-curves of the planet which are dominated by the rotation of spot 2015-A. Fig. 11 shows examples of images gathered in both periods displaying the bright feature in different wavelengths. Table 4 summarizes the observations obtained with large telescopes over 2015 and used in this study.

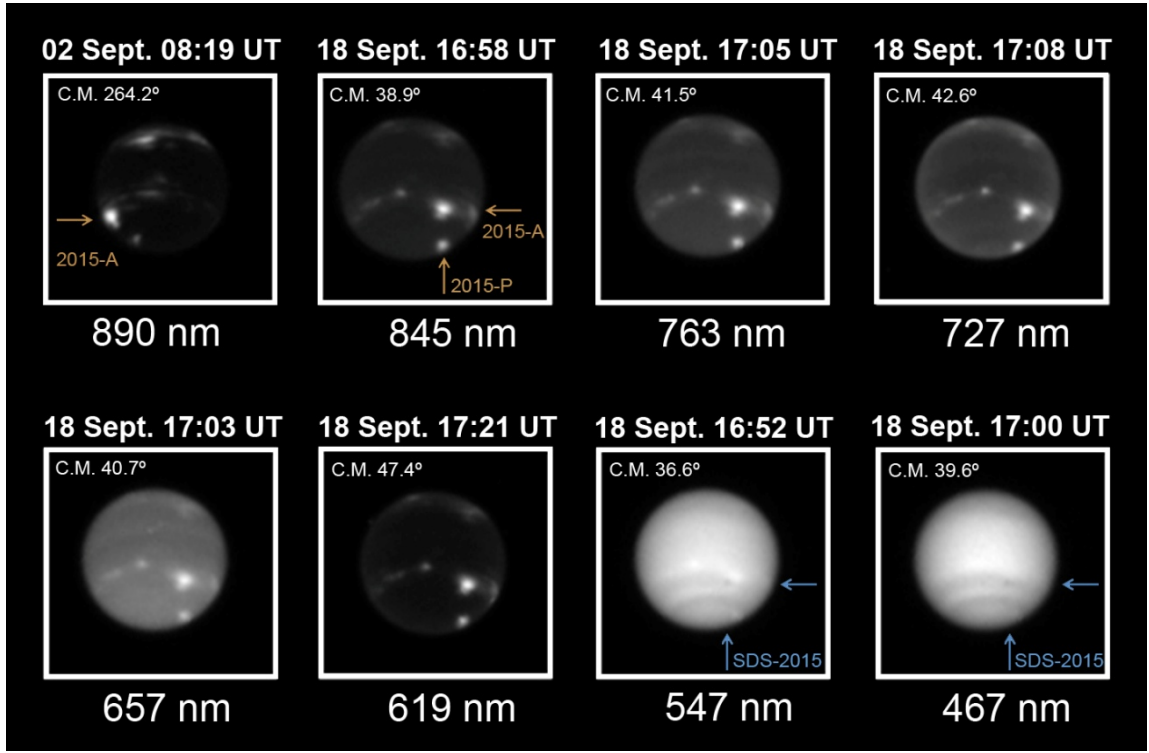

**Figure 11: HST observations of Neptune in September 2015.** Images acquired in wavelengths with strong methane absorption (890, 845, 727 and 619 nm) show bright cloud features with higher contrast than images in short wavelengths without methane absorption (657, 547 and 467 nm). North is up and West to the left. Observations on 2 September did not allow a good viewing angle of this atmospheric feature. Labeled features are commented in the text.

**Table 4: Neptune observations in 2015 with large telescopes.**

| Date | Telescope | Observer or PI | Instrument | Filters | Observing technique |
|---|---|---|---|---|---|
| 2015-01-10 | Keck II (10.0 m) | I. de Pater | NIRC2 | H | AO |
| 2015-07-13 | Calar Alto (2.2 m) | R. Hueso | PlanetCam | R, I, M1, M2, M3, RG1000, J, H | Lucky imaging |
| 2015-07-14 | Calar Alto (2.2 m) | R. Hueso | PlanetCam | R, I, M1, M2, M3, RG1000, J, H | Lucky imaging |
| 2015-07-25 | Keck II (10.0 m) | C. Baranec | NIRC2 | H | AO |
| 2015-07-28 | Palomar Hale (5.1 m) | S. H. Luszcz-Cook | P1640 | JH | AO |
| 2015-07-29 | Palomar Hale (5.1 m) | S. H. Luszcz-Cook | P1640 | JH | AO |
| 2015-08-05 | Keck II (10.0 m) | C. Baranec | NIRC2 | H, Kp | AO |
| 2015-08-10 | Lick Shane (3.0 m) | K. de Kleer | ShARCS | H, Ks | AO |
| 2015-08-29 | Keck II (10.0 m) | I. de Pater | NIRC2 | H | AO |
| 2015-08-30 | Keck II (10.0 m) | I. de Pater | NIRC2 | H | AO |
| 2015-08-31 | Lick Shane (3.0 m) | K. de Kleer | ShARCS | H, Ks | AO |
| 2015-09-02 | HST (2.4 m) | I. de Pater | WFC3 | 336, 467, 547, 619, 631, 727, 763, 750, 845, 889, 937, 953 | Image |
| 2015-09-03 | Lick Shane (3.0 m) | K. de Kleer | ShARCS | H, Ks | AO |
| 2015-09-04 | Lick Shane (3.0 m) | K. de Kleer | ShARCS | H, Ks | AO |
| 2015-09-18 | HST (2.4 m) | A. Simon | WFC3 | 467, 547, 619, 657, 727, 763, 845, 890 | Image |
| 2015-10-28 | Calar Alto (2.2 m) | A. Sánchez-Lavega | PlanetCam | R, I, M1, M2, M3, RG1000, J, H | Lucky imaging |

## 3. Analysis Methods

### 3.1. Image navigation and cylindrical projections

All images were navigated with the WinJupos free software (http://jupos.org/gh/download.htm). This software contains ephemeris of Solar System planets based on the semi-analytic VSOP87 (*Variations Séculaires des Orbites Planétaires*) description of their orbits (Bretagnon and Francou, 1988). The ephemeris system in WinJupos and predictions of Triton's position on different dates were compared with the Neptune Viewer tool in the Rings Node of NASA's Planetary Data System (http://pds-rings.seti.org) finding only minor differences on the order of 0.1° in planetary longitudes between both ephemeris calculations. Longitudes are measured with respect to Neptune's internal rotation period from the rotation of its magnetic field (System III) with period 16h 6m 36s (Archinal et al., 2011).

Navigation of amateur images of Neptune is challenging because diffraction and atmospheric seeing distorts the light from the planet resulting in a blurred disc without a well-defined limb. Additionally, the planet contains only a few atmospheric features and no visible information about its orientation in space. WinJupos was updated for this campaign to allow measuring positions over the Neptune disk by placing a grid over the planet that can be oriented and sized using the position of Triton in the same image. Ephemeris of Triton in WinJupos come from Seidelman (1992). Using Triton as a tie-point allows measuring longitudes and latitudes over most amateur images with an estimated mean error for mid South latitudes of 10° in longitude (for features at the central meridian) to 30º (for features separated 90° from the central meridian). Errors in latitude are typically 5°. The position of Triton was also used to navigate Pic du Midi, Robo-AO, Calar Alto and most HST images but it was generally not available in Palomar, Lick or Keck images which were navigated attending to the planet limb and the geometry of the cloud systems in the images. Cylindrical maps of these images produced with WinJupos were used to check the image navigation by testing the horizontal alignment in the maps of the zonal structure of the southern mid-latitude belt. We also used WinJupos to combine the data in the best images building nearly full maps of the planets in a few observation sets. For instance, the bright feature map in Fig. 9 from Palomar Hale was computed by combining two maps obtained when the feature was placed in the two limbs of the planet. All amateur images were processed using a variety of wavelet, high-pass and deconvolution filters by their authors. Images obtained at Pic du Midi were also processed using wavelets. All other images were left unprocessed except for adjustments of the image contrast.

**4. Feature identification, drift rates and sizes**

4.1. Analysis of 2013 data

A first analysis suggested that the entire set of amateur and Pic du Midi images from 2013 (Fig. 1) showed the same bright feature at middle south latitudes (Delcroix et al. 2014). This interpretation is more difficult to maintain when considering the variety of cloud systems captured in J (1.25 μm) and H (1.65 μm) bands with VLT/SINFONI from October 2013 and Keck observations from July 2013 (Fig. 2). We interpret the mid-latitude discrete features in these observations as two different bright features at close latitudes separated by 140° in longitude in July 2013 and simultaneously observed in Calar Alto observations close to the limb at a relative distance of 135° in longitude after traveling a relative distance of 85° that made them initially separate and then get closer. Both features were also observed in VLT images (Fig. 2).

Fig. 12 displays the longitude versus time positions of these features for 2013. This and later figures use an "extended longitude" system $L_{ext.}$ in which we add enough full rotations over the longitude system to match the data with straight lines. To get the actual longitude $L$ of an individual measurement shown in the figure it is enough to compute the modulo operation: $L = L_{ext}$ modulo 360°. The difficulty here is to know how many times the feature has drifted a full span of 360° over the longitudinal system of the planet. This is particularly difficult when considering observations separated from the others by long time intervals. In order to do that, linear fits to the data need to be calculated considering different multiple integers of 360° minimizing the residual longitudes with respect to the fits. At least three points are needed. Two fits, i.e., two atmospheric features (2013-A and 2013-B), represent an appropriate interpretation of the data that agrees with the two bright features in VLT images and the two features observed in Calar Alto a few days earlier. The feature initially detected in Pic du Midi (2013-A) seems to have survived from its first detection on 27 June to the VLT observations in 10 October. The bright feature observed in Keck II data in 31 July seems to be a different feature (2013-B) that can be tracked to VLT data from October and that also fits some of the amateur detections. These fits also match relatively well the amateur observations of bright features in the planet. Latitudes and drift rates of these two features, calculated without considering the amateur observations, are 19.0±0.3°/day (westward) for 2013-A at planetographic latitude -40±7° and 19.9±0.1°/day (westward) for 2013-B at planetographic latitude -46.7±3.5°.

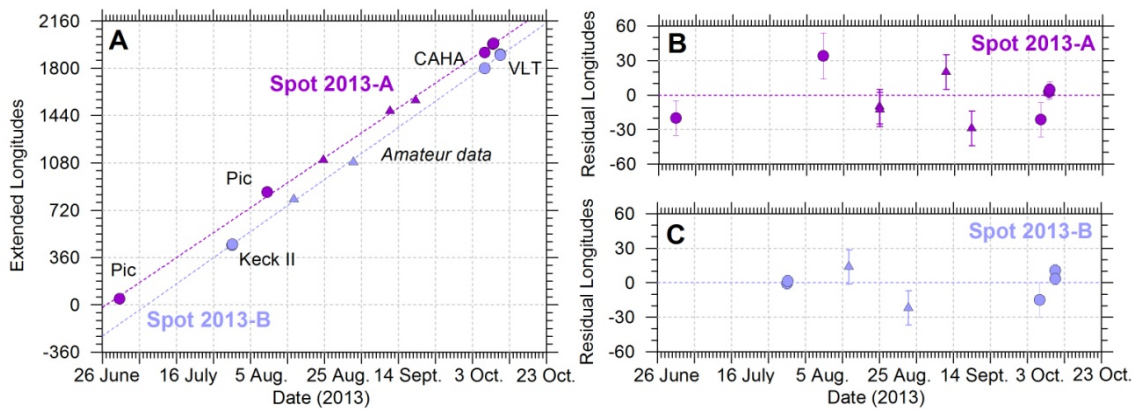

**Figure 12: Bright features tracked in 2013.** (A): Longitudes of all mid-latitude bright features as a function of time. Extended longitudes are shown correcting for the number of feature rotations around the planet. Measurements from large telescopes are marked as large dots and amateur observations are shown with triangles. (B) and (C): Residuals in longitude after subtracting the actual data from the linear fits. Individual error bars are approximate and could be larger in the amateur data.

Unfortunately, this fit is not unique and can be interpreted as a best guest that agrees with the data. A second alternative exists based on adding a different number of rotations to the positions of the features as time passes. The alternative model is presented in Fig. 13 and results in drift rates of these two features, now named 2013-A* and 2013-B* to indicate their different and much faster drift rate of 36.9±0.4°/day (westward) for 2013-A* and 40.1±0.2°/day (westward) for 2013-B*. Again, these drift rates were calculated without considering the amateur observations. These drift rates would result in both features colliding around November 2013.

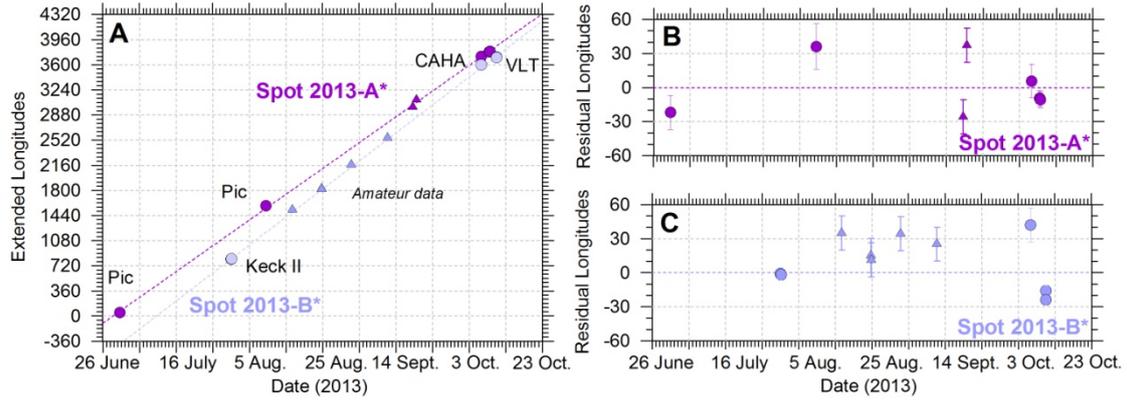

**Figure 13: Alternative fit to bright features in 2013.** As in Figure 12 but for the alternative fits described in the text.

Deciding between both alternative models is not easy. If we define a parameter $D$ as the average of the absolute value of the differences between the measurements and the linear fits as: $D = \sqrt{\chi^2/N}$, where, $\chi^2$ is the sum of the squared differences of all longitudinal measurements and the fit, and $N$ is the number of points, then the value of $D$ for the first model with slow drift rates is $D=9.0°$ for spot 2013-A and $D=8.9°$ for spot 2013-B. Values of $D$ for the second model are $D=9.0°$ for 2013-A* and $D=10.2°$ for 2013-B*. These numbers show a slightly poorer fit in fitting the data and correspond to the data from large telescopes only. If we also consider the amateur data in Figs. 12 and 13 the first case also results in a better overall fit of the amateur data ($D=7.7°$) than the fits with fast drift rates in Fig. 13 ($D=10.6°$). We present further arguments in favor of the first set of fits in section 5 when we compare the drift rates with zonal winds.

4.2. Analysis of 2014 data

Keck images in Fig. 4 show the simultaneous presence of two bright features at nearby latitudes but separated in longitude (2014-A and 2014-B). All other observations in 2014 show only one of these bright features and the large gaps in the temporal sampling make the identification of bright features with features 2014-A or 2014-B in Fig. 4 not straightforward. We locate the spot longitudes using an extended longitude system in which we add a full 360 deg approximately every 18 days and we fit each observation to one of the features. As in the previous case different alternative models are possible. The simplest model is shown in Fig. 14 and results in spot 2014-A having a planetographic latitude of -36.7±2.7° and westward drift rate of 19.5±0.3°/day with spot 2014-B having a planetographic latitude of -38.4±3.0° and westward drift rate of 20.2±0.3°/day. Fig. 14 also shows that spots 2014-A and 2014-B could have merged around mid October 2014 yielding a single bright feature drifting westward at 22.7°/day over that period of time.

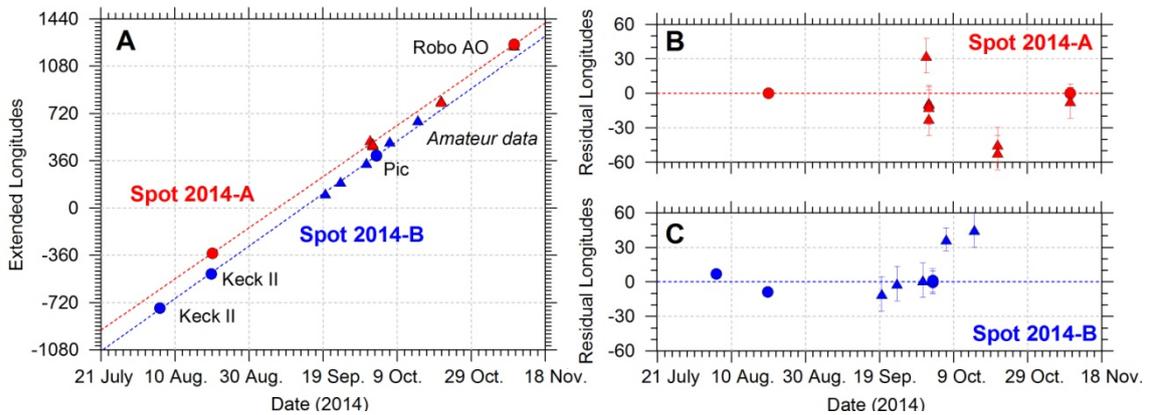

**Figure 14**: **Bright features tracked in 2014.** (A): Longitudes of all bright features as a function of time. Extended longitudes are shown correcting for the number of feature rotations around the planet. Measurements from large telescopes are marked as large dots and amateur observations are showed with triangles. (B) and (C): Residual longitudes after subtracting the linear fits calculated from data obtained in large telescopes only. Individual error bars are approximate and could be much larger in the amateur data. Both spots are identified by a color code.

Again, the fit is not unique and an alternative set of fits with a fast drift rate for both features is possible. This alternative model is also calculated without considering the amateur data and is shown in Fig. 15. The alternative fits result in the same latitudes for both features and drift rates of 46.0±0.4°/day (westward) for 2014-A* and 44.5±0.2°/day (westward) for 2014-B*. These drift rates would result in both features colliding around November 2014, shortly before the January 2015 observations where a full map of the planet shows only one single spot (Simon et al., 2016).

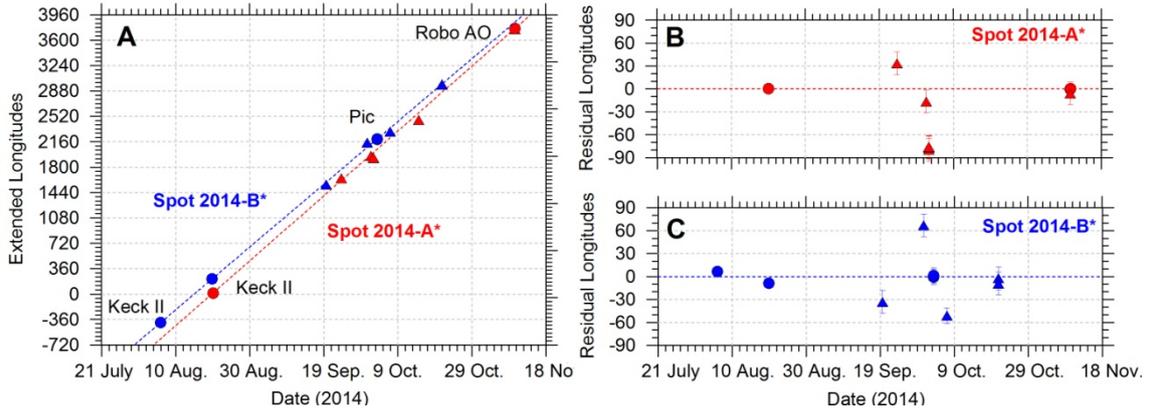

**Figure 15: Alternative fit to bright features in 2014.** As in Figure 14 but for the alternative fits described in the text.

The differences between the different fits and the spots positions result in values of *D* without considering amateur data for the first model with slow drift rates of *D*=0.0° for spot 2014-A and *D*=2.2° for spot 2014-B. Values of this parameter for the second model are *D*=0.0° for spot 2014-A* and *D*=1.9° for spot 2014-B*. When we consider the amateur data, the first set of fits (2014-A and 2014-B) gives *D*=7.2° while the second set of fits (2014-A* and 2014-B*) gives a value significantly higher, *D*=14.4°. However, we will show in section 5 that a comparison of the drift rates with zonal winds favors this second set of fits (2014-A* and 2014-B*).

4.3. Drift rates of bright features in 2015

Calar Alto (Figs. 5 and 6), Keck (Fig. 10), and HST (Fig. 11) observations show that a single bright feature was present in Neptune's mid southern latitudes from July to September 2015. A nearly full map of the planet obtained in January 2015 from Keck images also shows only one bright feature (Fig. 3 in Simon et al., 2016). Using data from July to November from large telescopes together with a selection of outstanding amateur images in November 2015 allows to obtain an unambiguous fit to the data with a westward drift rate of 24.48±0.03°/day for the bright feature (spot 2015-A) (Fig. 16A). The fit is unique due to the high number of observations and short time difference between many of them and the small error in the drift rate comes from the long time span of the observations (153 days). The planetographic latitude of the feature from the ensemble of measurements gives a value of -40.1±1.6° (Fig. 16B). Sussenbach et al. (2017) present a similar analysis for a reduced selection of 14 amateur images from two observers finding a westward drift rate of 24.0°/day. They also present an analysis of the contrast of the feature in amateur images and strategies to maximize the contrast of Neptune features with amateur equipment.

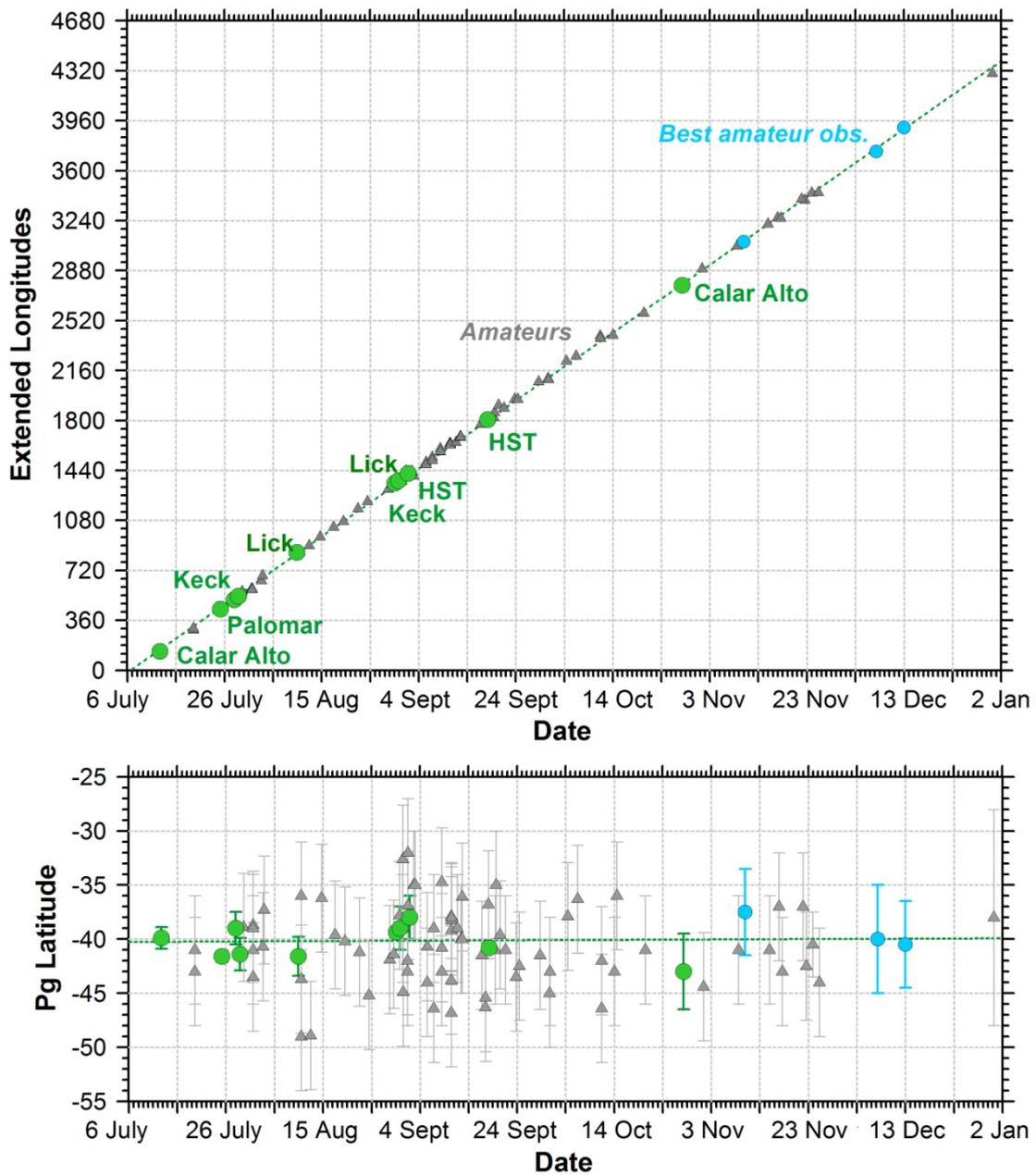

**Figure 16: Global tracking of spot 2015-A.** (A): Longitudes of all bright features as a function of time. Extended longitudes are shown correcting for the number of feature rotations around the planet. Measurements from large telescopes are marked as large green circles and amateur observations are showed with triangles. Amateur observations of outstanding quality selected after the last observations with large telescopes are shown with blue circles and are also used for the longitude time fit. (B): Planetographic latitudes of the centroid of the bright feature.

Residual longitudes calculated from subtracting the linear fit from the extended longitudes are shown in Fig. 17. The data shows that spot 2015-A did not move with a constant drift rate over its life. A possibility to describe the variations of its drift rate is by fitting a sinusoidal function to the residual longitudes. This would represent an oscillation in longitude with amplitude of 16.0±2.5° and a period of 90±3 days. Uncertainties in these numbers are estimated from a full examination of the space of parameters defining the fit (amplitude, period and phase) and from comparisons with an equivalent analysis of only the amateur data. An analysis of the amateur data alone shows a similar fit with amplitude of 13.5±2.5° and a period of 95±5 days. This kind of longitudinal oscillation has been found previously in dark and bright features in Neptune. The most well-known are the features observed by Voyager 2 (Smith et al., 1989) with oscillation of the Great Dark Spot (GDS), Dark Spot 2 (DS2) and the bright Scooter. The Dark spots had longitudinal oscillations accompanied by significant changes in latitude, but the Scooter had an oscillation of longitude of 20° with a period of 120 days and only a very small (< 0.5 deg) oscillation in latitude (Sromovsky et al., 1993; Sromovsky et al., 1995). Other oscillations in Neptune dark cloud features were found during 1991 to 2000 (Sromovsky et al. 2002). For spot 2015-A, since only one full oscillation is detected, alternative possibilities could exist. For instance, a parabolic fit to the global longitudinal trend that could correspond to a small change of latitude and drift rate would largely remove the oscillation. Therefore, the longitudinal oscillation, though similar to previous detections of oscillations in Neptune's atmosphere, is only one possibility.

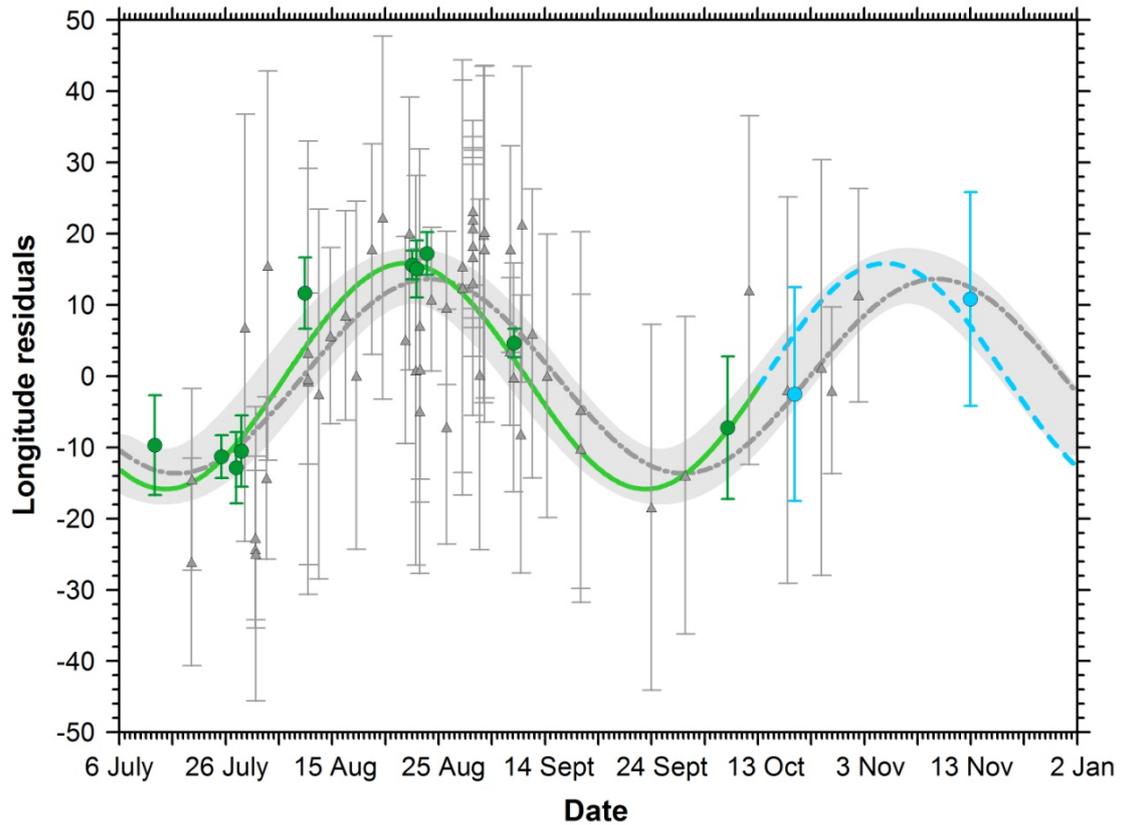

**Figure 17: Residual longitudes of spot 2015-A calculated from subtracting the linear fit from the longitude of the feature**. Green circles correspond to data from large telescopes. Blue circles correspond to outstanding amateur images in the time period not covered by observations with large telescopes. Grey triangles represent amateur data. Amateur data with estimated uncertainties in longitude larger than 30° have not been taken into account. A sinusoidal fit to the data from large telescopes only is shown in green and extended in blue in the area covered only by amateur observations. A sinusoidal fit to the amateur data alone is shown with a grey dashed line. The range of sinusoidal models that fit the combination of all observations is shown with a grey shaded region. The nature of this oscillation largely depends on the linear fit to the data and other possibilities exist.

4.4. Size measurements of the mid-latitude bright features

Figs. 2 to 11 show that the bright spots at mid-latitudes appear with a variety of sizes on different images being the smallest structure the one observed in 2013-B. We measured the size of the brightest area of these spots using images where the feature appears not saturated and a true measurement of its size can be obtained. In saturated images the apparent sizes can increase by a factor of two. Simultaneous observations in H (1.65 μm), Kp (2.12 μm) and Ks (2.15 μm) bands when available, show very similar structures and essentially the same sizes for the different spots at these long wavelengths. The measurements are presented on Table 5. Each measurement corresponds to four measurements of the size of the feature either over the same images or over close in time images when available. Measurement errors are computed from the standard deviation of the ensemble of measurements and do not contain the limited size of the telescope PSF which is shown in an additional column and is generally bigger. We adopt PSF as the metric that determines measurement errors in the size of these features. We could not measure the size of spot 2013-A due to the lack of a non saturated image with enough spatial resolution. Spot 2013-B has a mean longitudinal size of 4,600±900 km and a mean latitudinal size of 3,000±900 km. Spot 2014-A is just slightly larger and spot 2014-B is about 30% longer and 50% wider than 2013-B. The bright spot 2015-A is the largest of these systems and has a mean longitudinal size of 8,000±900 km and a mean latitudinal size of 6,500±900 km. This is ~3.8 times the area of spot 2013-B, the smallest one here analyzed, and has approximately the same area (10% larger) as the combination of areas of spots 2014-A and 2014-B.

For spot 2015-A HST observations also allow to measure its size in visible wavelengths. At shorter wavelengths the feature loses contrast and size. Its size at red-wavelengths, where it is still easily observable, is 5,100±1400 km in longitude with a mean latitudinal size of 4,500 ± 1400 km. It is not possible to distinguish this feature at shorter wavelengths.

**Table 5: Sizes of mid-latitude bright features**

| Date & Feature | Telescope | Wavelength or band | Longitudinal size (km) | Latitudinal size (km) | dLon* (km) | dLat* (km) | PSF (km) |
|---|---|---|---|---|---|---|---|
| **2013-B** | | | | | | | |
| 2013-07-31 | Keck II | H | 4,600 | 3,000 | 450 | 350 | 900 |
| **2014-A** | | | | | | | |
| 2014-08-20 | Keck II | H | 5,300 | 3,800 | 400 | 400 | 900 |
| **2014-B** | | | | | | | |
| 2014-08-06 | Keck II | H | 5,600 | 4,500 | 800 | 800 | 900 |
| 2014-08-20 | Keck II | H | 6,300 | 4,600 | 800 | 500 | 900 |
| **2015-A** | | | | | | | |
| 2015-07-13 | Calar Alto | H | 6,900 | 6,800 | 1,200 | 1,000 | 4,000 |
| 2015-07-25 | Keck II | H | 7,500 | 5,300 | 500 | 300 | 900 |
| 2015-07-28 | Hale | J, H | 9,200 | 7,200 | 900 | 700 | 1,500 |
| 2015-08-20 | Keck II | H | 8,400 | 7,200 | 1,800 | 300 | 900 |
| 2015-08-31 | Shane | H | 8,000 | 5,800 | 700 | 900 | 900 |
| **2015-A (mean size in H band):** | | | 8,000 | 6,500 | 900 | 600 | 900 |
| 2015-09-18 | HST | 845 nm | 6,300 | 4,900 | 700 | 600 | 1,700 |
| 2015-09-18 | HST | 657 nm | 5,100 | 4,500 | 400 | 400 | 1,400 |

(*) dLon and dLat correspond to the statistical error from a set of measurements of the size of the different bright features.

4.5. Neptune's Southern Dark Spot

HST images from the OPAL program (Simon et al., 2016) at short visible wavelengths, 467 and 547 nm, show that the bright spot 2015-A is accompanied by a dark feature centered at -45° latitude that separates the north and south branches of the bright cloud (Fig. 11). Fig. 18 shows an enhanced view of a color composition of 2015 HST images of Neptune where the dark spot is more apparent. Photometric scans of the dark spot show that it is 7% darker at 467 nm than its environment and extends about 25±5° in longitude or 7,500±1,500 km and 3,000±1,000 km in latitude.

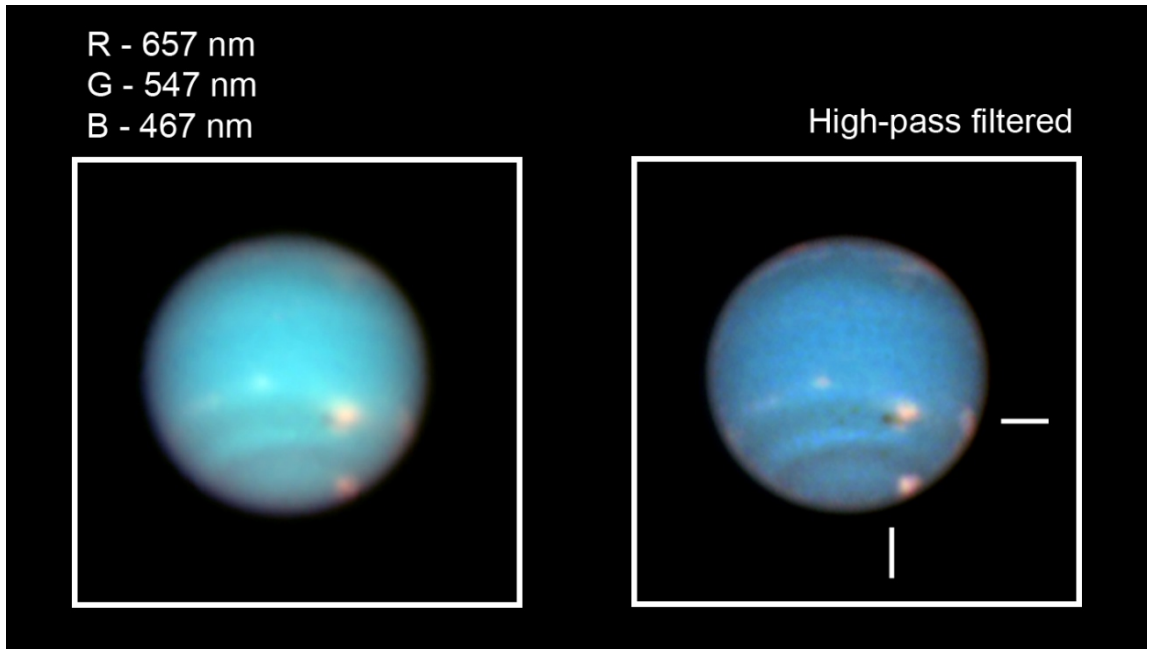

**Figure 18: Neptune's Southern Dark Spot (SDS-2015) in September 2015.** Note that the dark spot marks the Western separation of the north and south branches of the bright companion clouds. This same double feature is present at least from January 2015 and is similar to the morphology of spot 2014-A in Fig. 4 observed in August 2014.

Subsequent optical-wavelength HST observations in 2016 (Wong et al., 2016) confirmed the presence of a dark spot at latitude -46°, making this the fifth dark spot ever seen on Neptune. Previous dark spots are the Great Dark Spot discovered by Voyager 2 in 1989 (GDS-89), a secondary smaller dark spot (DS2) also imaged by Voyager 2 (Smith et al. 1989), a north-hemisphere dark spot discovered in 1994 (NDS-1994) in HST images (Hammel et al., 1995), and another northern dark spot discovered in 1996 (NDS-1996; Sromovsky et al., 2001b, 2002). We call the dark spot in 2015 SDS-2015, which stands for "Southern Dark Spot" discovered in 2015.

Similarly to previous dark features in Neptune, this dark spot could be a long-lived dark vortex with cloud tops located at about 1 bar in blue images and deeper than the bright clouds visible as spot 2015-A. Neptune's dark spots have been diverse in terms of size, aspect ratio and latitudinal drift. In most cases, they have oscillations in longitude and have accompanying bright cloud systems with different cloud morphologies. The largest vortex observed on Neptune, the GDS-89, was the only feature seen to drift equatorward and dissipate. This behavior was explained by LeBeau and Dowling (1998) as a response to the weak meridional shear of the zonal wind. Stratman et al. (2001) simulated the Great Dark Spot and explained the formation of its bright companion as an orography-like cloud formed by the perturbation in zonal wind streamlines imposed by the anticyclone. An interesting aspect of the GDS bright companion clouds and the simulations is the asymmetry in the location of the bright companion clouds with respect to the vortex which for the GDS-89 were rimming the poleward edge. In numerical simulations by Stratman et al. (2001) the asymmetry is produced by the interaction of the anticyclonic vortex with the zonal wind resulting in a lower level of wind turbulence in the poleward edge of the vortex and more constant cloud production. In the SDS-2015 the bright companion cloud is significantly different, being mainly located in the northern part of the vortex and making an extremely large cloud system (spot 2015-A) with a well-defined shape preserved from January 2015 until at least September 2015. A similar shape is also found in the smaller spot 2014-A in August 2014.

4.6. Extended analysis of the bright features from 2013 to 2015

The southern mid-latitude bright features from 2013 to 2015 are sufficiently close in terms of their latitude that one could wonder about the possible relation among them. Fig. 19 presents the latitudinal behavior of these bright cloud systems. If we consider linear trends to the latitudinal data, then spots 2013-B and 2014-B show latitudinal drifts of +0.022°/day and +0.025°/day respectively. In fact, the latitudinal fit to the 2013-B alone and the latitudinal fit to the 2014-B data alone are very similar suggesting the possibility that spot 2013-B migrated in latitude to become spot 2014-B. The latitudinal drift rate of spot 2013-A is not well constrained by the data and prevents an assessment of whether spot 2013-A and 2014-A could be the same or different structures.

The latitudinal trends in Fig. 19 for 2014 suggest that features 2014-A and 2014-B could have reached the same latitude around October-November 2014 which is similar to the results of longitudinal drift rates presented in section 4.2. In particular, we recall that the fast drift rates in the alternative fits 2014-A* and 2014-B* predict this encounter to have occurred in November 2014, while fits 2014-A and 2014-B would have resulted in a merger of both features in March 2015, which is too late to explain the isolated spot 2015-A in January 2015.

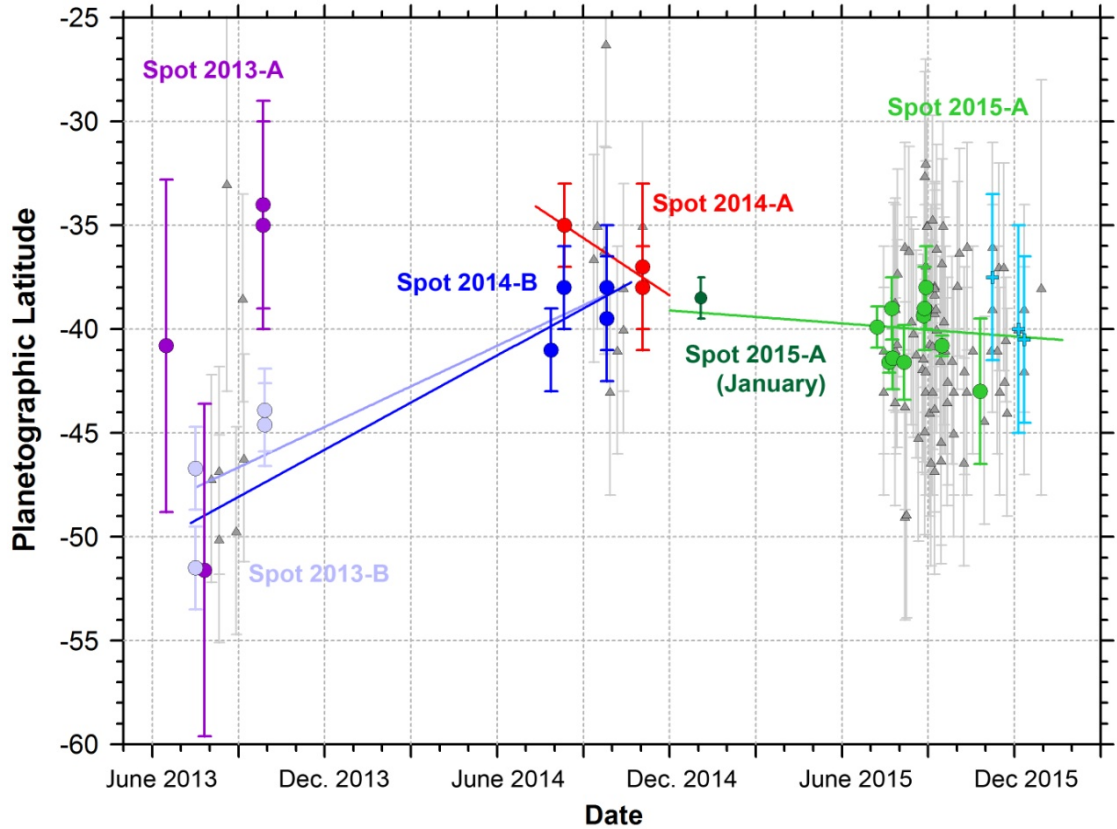

**Figure 19: Planetographic latitudes of bright features in 2013 to 2015.** Circles represent latitudes from images acquired with large telescopes, triangles come from amateur observations, crosses correspond to outstanding amateur observations in 2015. Colored linear fits to the 2013-B (clear blue), 2014-A (red), 2014-B (dark blue) and 2015-A (green) data are shown. Amateur data is not taken into account for these fits.

We note that a combined longitudinal analysis of the 2013, 2014 and 2015 features is not possible, since the features moved in latitude changing their longitudinal drift rates in a way that would introduce new uncertainties.

4.7. Tracking of other cloud systems

Many of these observations also show other bright features at different positions in the planet (Fig. 20). The most conspicuous is a feature at the northern limb corresponding to North tropical latitudes and here called 2015-N. This feature was tracked from 14 July to 28 October on images from different telescopes. It was also visible in amateur observations in at least 12 amateur images from 8 different observers over 106 days. This results in a drift rate of +71.0±0.3°/day and planetographic latitude of 23.9±3.9° from the analysis of data from large telescopes only (Fig. 21, panels A-C). The last observation of this feature was excluded from the analysis as it was acquired with the feature close to the limb and under bad observing conditions.

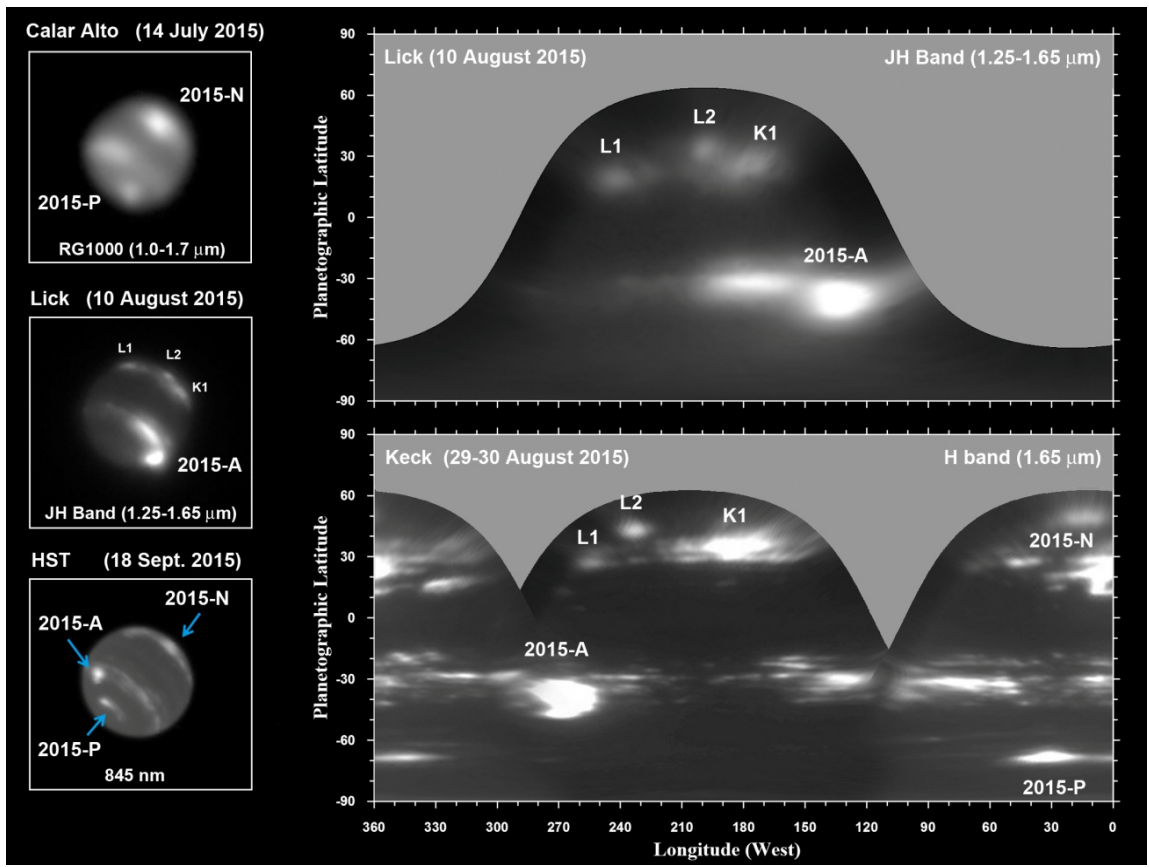

**Figure 20: Identification of long-lived cloud features on Neptune in 2015.** Images on different dates show a variety of cloud features with different contrast depending on wavelength. A North feature at tropical latitudes in the limb (2015-N) was observed repeatedly from July to October 2015, while other bright features at close north latitudes were only observed twice in August 2015 (features L1, L2 and K1). A south subpolar feature (2015-P) was also identified from July to September 2015.

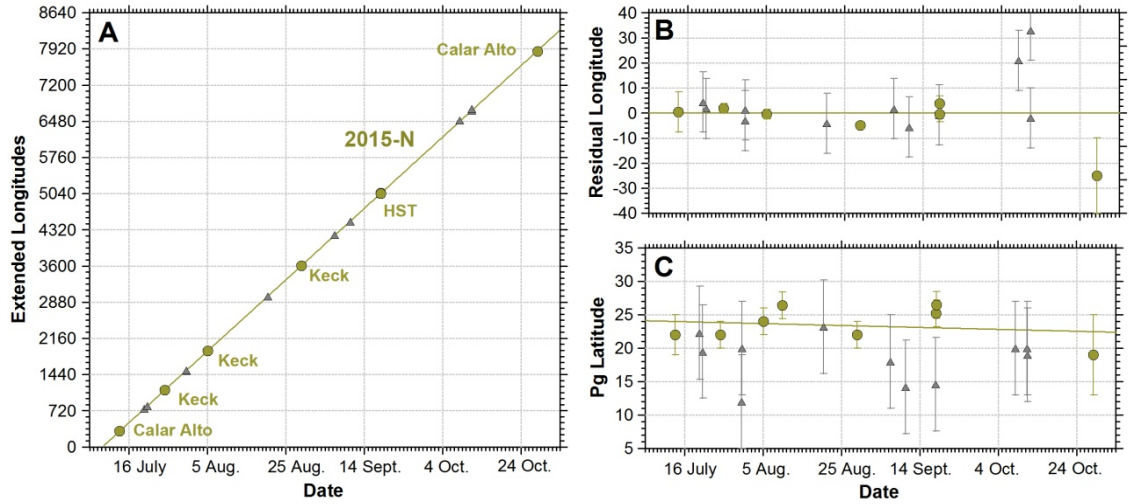

**Figure 21: Global tracking of feature 2015-N.** (A) Longitudinal positions from large telescopes are shown with filled circles, amateur observations are showed with triangles. (B) Residual longitudes of 2015-N after subtracting the linear fit to the data from large telescopes only. (C) Planetographic latitudes of the centroid of 2015-N. Fits to the data do not consider the last Calar Alto observation.

Many of the high resolution images in H and K bands or in PlanetCam filters J and H present several large cloud systems that could not be identified repeatedly on different images. Only three stable systems, K1, L1 and L2 were observed in August 2015 that could be identified in two different dates and they are also labeled in Fig. 20. Their drift rates and planetographic latitudes correspond to 54.0±1.0°/day and 29.6±4.5° for K1, 54.3±0.6°/day and 25.5±3.5° for L1, 37.6±0.8°/day and 39±5° for L2.

Additionally, a polar feature at latitude -69.2±3.5°, here called 2015-P, was observable on 6 dates from 14 July to 18 September and was visible in at least two amateur observations over about 67 days. This feature could be related to the South Polar Feature (SPF) initially detected in Voyager 2 images (Smith et al., 1989) showing a large dispersion of wind measurements (Limaye and Sromovsky, 1991) and later in HST observations (Sromovsky et al., 1993; Karkoschka, 2011). Sromovsky et al. (1993) showed that the SPF was composed of high clouds drifting fast but the SPF as a whole was drifting slow at a rate of 15.968±0.004 hr (-4.77±0.14 deg/day). This period was later improved to 15.96628 ±0.00005 hr (-4.8276±0.0017 deg/day) by Karkoschka (2011) using a combination of Voyager 2 and HST observations. This feature could also be related to the deep sub polar features observed in 2009 at pressure levels higher than 1.25 bar (Irwin et al., 2011) that were absent in 2013 on VLT/SINFONI observations that showed only shallow sub polar clouds (Irwin et al., 2016). Therefore, considerable variability has happened in this region of the planet in recent years. The tracking of this 2015-P polar feature from July to October 2015 is not simple due to its elongated structure and the evolution of features in the highest spatial resolution data (HST and Keck).

Figure 22 shows two trackings for this polar feature. On the one hand, Figure 22A shows the long-term tracking of the center of this feature. Its drift rate is -4.6±0.5 deg/day (with a period of 15.973±0.015 hr) but longitudinal residuals from the fit (Fig. 22B) show large dispersions that could be associated to the elongated shape of the structure, its fast evolution in time, and the viewing angles sometimes close to the limb. Figure 22C shows the latitudinal position of the 2015-P structure. These data could be interpreted as a detection of the classical SPF with a compatible drift rate and period to published values. On the second hand, HST observations in 18 September show a compact bright cloud, here called 2015-P1* as part of the 2015-P system. Tracking this bright core over 4.8 hours (Fig. 22D) results in a fast rotation period of -160±30°/day. The bright cloud is shown in panel Fig. 22E. After a single rotation its structure change considerably (Fig. 22F) and the tracking of the brightest point in 2015-P, a feature here called 2015-P2* resulted in a different rotation period of -112±35°/day. This second feature can be tracked with less confidence due to its particular position over the disk in the images. Both values are compatible within the retrieved errors with the zonal winds at polar latitudes.

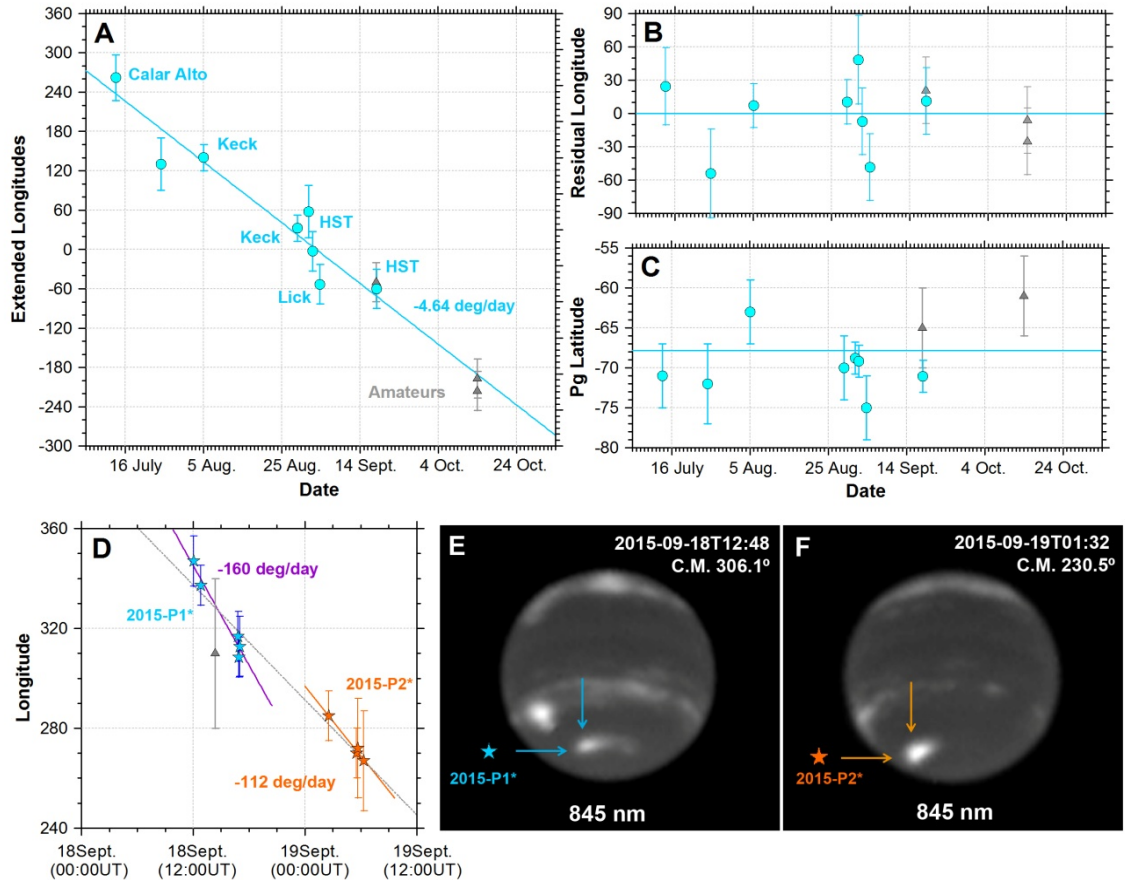

**Figure 22: Global tracking of South polar features.** (A) Longitudinal positions of 2015-P. Filled circles show observations with large telescopes, triangles show amateur observations. (B) Residual longitudes of 2015-P after subtracting the linear fit to the data. (C) Planetographic latitudes of the centroid of 2015-P. (D) Longitudinal positions and fits to compact bright features in HST observations in September 2015. (E) One of the observations of the bright compact source here called 2015-P1* tracked in (D) with blue stars. (F) Second set of observations with a bright source with a different morphology, here called 2015-P2*, and tracked in (D) with orange stars.

## 5. Zonal-wind velocities

Fig. 23 translates the drift rates in values of zonal winds and compares those values with Neptune zonal winds from Voyager 2 (Limaye and Sromovsky, 1991; Sromovsky et al. 1993; for a recent updated review of Neptune's zonal-wind profile see Sánchez-Lavega et al., 2017). Features 2015-N, 2015-P, K1, L1 and L2 have motions that follow closely the wind measurements obtained at the time of Voyager 2. This overall agreement is probably due to the long-term survival of the features tracked, which minimizes errors due to longitude measurement. The average of features 2015-P1* and 2015-P2* matches the zonal wind profile.

Most of the long-lived features at mid-latitudes from the standard fits in Fig. 12 (2013 data), Fig. 14 (2014 data) and Fig. 16 (2015 data) do not match the wind profile. Spot 2013-B is the only one that truly fits Voyager zonal winds while spot 2013-A marginally fits the Voyager wind profile if we consider the large uncertainty in its latitudinal position. If we consider the alternative fits for 2013 data appearing in Fig. 13, faster zonal speeds are obtained. Then, feature 2013-A* (the symbol * only indicates that its zonal speed corresponds to the alternative fits to the longitudinal trends) fits the Voyager zonal-wind profile but not the feature 2013-B*. We must stress that the alternative fits come in pair, so that either we have the zonal speeds of features 2013-A and 2013-B, or those of 2013-A* and 2013-B*. Therefore, the 2013 data is overall more compatible with Voyager zonal winds when considering the standard fits (2013-A and 2013-B) to the longitudinal trends.

Standard fits to the 2014 data for features 2014-A and 2014-B do not match the Voyager zonal wind profile. In this case, alternative fits (2014-A* and 2014-B*) match better the Voyager zonal wind profile and seem to be a better representation of the data from that point of view.

The drift rate of spot 2015-A was measured without ambiguities and results in a zonal speed that is separated 30 m/s with respect to the Voyager wind profile. This deviation is greater than typical wind variations in Neptune at south mid-latitudes when using long-term tracking (Limaye and Sromovsky ,1991; Sromovsky et al., 2001b; Fitzpatrick et al. 2014). If we consider that the bright cloud could be rooted to the SDS-2015, a different wind result can be obtained for this bright cloud. Using the latitude of SDS-2015 which is located southward at a latitude of -45.0±1.0°, and the zonal drift of the bright spot 2015-A the zonal speed for its drift rate translates into a zonal velocity -89.5 m/s that matches the Voyager mean zonal wind at its latitude. This seems a better assessment of the velocity of this extended dark spot plus bright companion system.

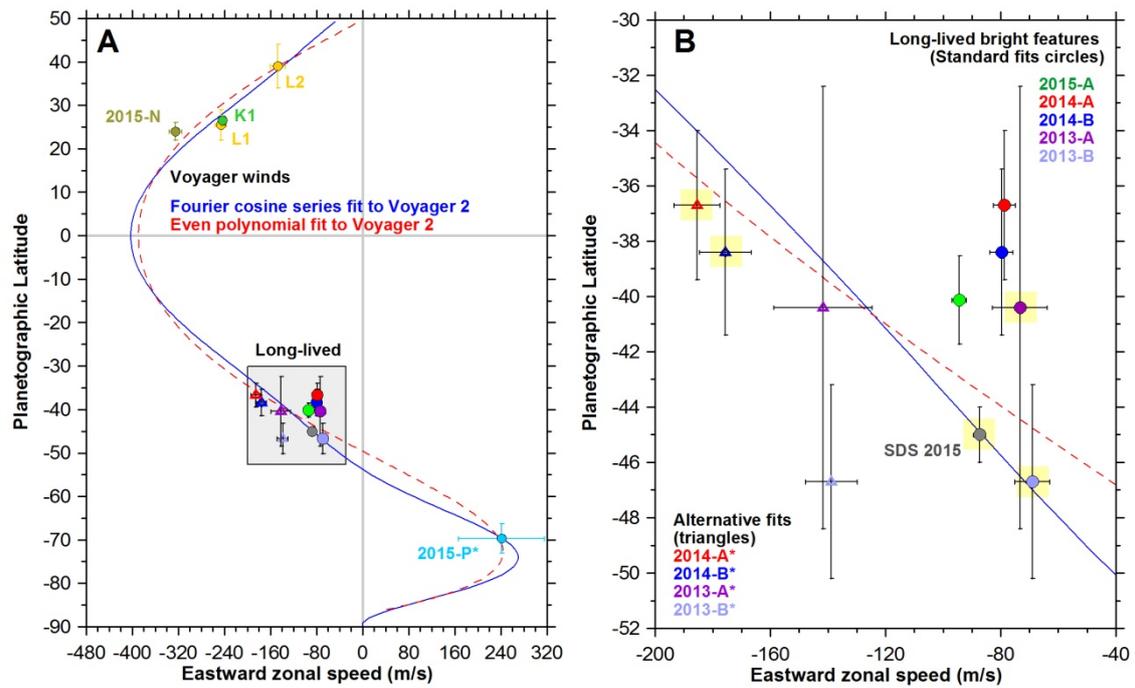

**Figure 23: Cloud feature motions in terms of zonal winds compared with cloud tracking in Voyager 2 images**. Voyager wind profiles. Blue solid line is a Fourier cosine series fit given in Sánchez-Lavega et al. (2017) to the data in Limaye and Sromovsky (1991) and Karkoschka et al. (2011). Red dashed line shows the 6$^{th}$ order even fit to the Voyager wind speeds given by Sromovsky et al. (1993). (A) All wind data from this work. Standard fits appear with colored solid circles. Alternative fits appear with open triangles. Zonal wind error bars are mainly due to the dispersion of latitude and the transformation of degrees per day to meters per second which is different at different latitudes. A single point is used to represent both features 2015-P1* and 2015-P2*. (B) Zoom over the long-lived features which produce ambiguous measurements. The speed associated with SDS 2015 (grey filled circle) takes into account the drift rate of its bright white companion (spot 2015-A) and the latitude of SDS 2015 and may be considered more accurate than the 2015-A measurement (green filled circle). Preferred solutions for the zonal motions of the different spots are highlighted with yellow boxes for each measurement.

The association between the dark vortex, SDS-2015, and bright spot 2015-A suggests that perhaps some of the bright spots seen in 2013–2015 are companion clouds formed near a dark vortex. In the absence of HST observations of Neptune in 2013 and 2014, we can only speculate if the differences of zonal winds from spots 2014-A and spot 2014-B could be explained by the presence of dark vortices at southern latitudes that would form these bright cloud systems as northern bright companions. However this remains highly speculative. Instead, features 2014-A and 2014-B could just move at the zonal speeds implied by the alternative fits 2014-A* and 2014-B*, which have a poorer fit to the scarce data in 2014. The motions of features 2013-A and 2013-B is compatible with Voyager zonal winds and does not need a deep root vortex explanation and deviations from the Voyager zonal wind are compatible with the uncertainties in latitude and can be also explained by the scarce data over 2013.

Table 6 presents a summary of the identified features over this work.

**Table 6: Summary of identified features**

| Feature | Interval | Ref. Time | Ref. Lon (°) | a (°) | b (°/day) | Latitude (°) | *Drift rate* (°/day) | u (m/s) |
|---|---|---|---|---|---|---|---|---|
| 2013-A | 2013-07-01 – 2013-10-10 | 2013-08-10T00:45 | 139 | -1350 | 18.99 | -40.4 ± 8.0 | 19.1 ± 0.2 | -73 ± 10 |
| 2013-B | 2013-07-31 – 2013-10-12 | 2013-07-31T15:07 | 101 | -1628 | 19.85 | -46.7 ± 3.5 | 19.9 ± 0.1 | -69 ± 6 |
| 2014-A* | 2014-08-20 – 2014-11-10 | 2014-08-20T13:23 | 15 | -4123 | 45.96 | -36.7 ± 2.7 | 46.0 ± 0.4 | -186 ± 8 |
| 2014-B* | 2014-08-06 – 2014-10-21 | 2014-08-20T13:56 | 218 | -3780 | 44.47 | -38.4 ± 3.0 | 44.5 ± 0.4 | -176 ± 9 |
| 2015-A | 2015-07-13 – 2015-12-13 | 2015-09-18T16:54 | 5.8 | -260 | 24.48 | -40.1 ± 1.6 | 24.48 ± 0.03 | -94 ± 3 |
| SDS-2015 | 2015-09-18 – 2015-09-19 | 2015-09-18T16:50 | 13.2 | --- | --- | -45.0 ± 1.0 | 24.48 ± 0.03 | -87 ± 2 |
| 2015-N | 2015-07-14 – 2015-10-28 | 2015-08-05T13:51 | 109 | -929 | 70.84 | 23.9 ± 3.9 | 71.0 ± 0 3 | -325 ± 11 |
| K1 | 2015-08-05 – 2015-08-30 | 2015-08-30T13:18 | 182 | -1894 | 54.05 | 26.5 ± 1.0 | 54.0 ± 1.0 | -243 ± 7 |
| L1 | 2015-08-10 – 2015-08-30 | 2015-08-30T13:18 | 254 | -2198 | 54.30 | 25.5 ± 3.5 | 54.3 ± 0.6 | -246 ± 10 |
| L2 | 2015-08-10 – 2015-08-30 | 2015-08-30T13:18 | 233 | -1494 | 37.60 | 39.0± 5.0 | 37.6 ± 0.8 | -147 ± 13 |
| 2015-P | 2015-07-14 – 2015-10-14 | 2015-08-29T12:29 | 33 | 319 | -4.6356 | -69.2 ± 3.5 | -4.6 ± 0.5 | --- |
| 2015-P1* | 2015-09-18T[12:02 – 16:58] | 2015-09-18T12:49 | 337 | 45768 | -159.94 | -67.4 ± 2.1 | -160 ± 30 | -312 ± 65 |
| 2015-P2* | 2015-09-19T[01:32 – 06:18] | 2015-09-19T05:32 | 285 | 32108 | -111.81 | -67.9± 2.0 | -112 ± 35 | -214 ± 70 |

**Note:** *a* and *b* give a linear fit to the tracking of the feature with Lon=*a*+*b**(Julian Date – JD0). Where JD0 = 2456400.0 for the data obtained in 2013, JD0=2456800.0 for the data obtained in 2014, and JD0=2457200.0 for the data obtained in 2015. The parameters *a* and *b* are given with enough accuracy to retrieve the longitude of the feature in the time interval when the feature was observed and do not represent the uncertainties over these values.

**6. Summary and Conclusions**

- Amateur observers using telescopes with 28- to 50-cm apertures have been able to repeatedly find bright features on Neptune at its Southern mid-latitudes in 2013, 2014 and 2015. Atmospheric features at other latitudes can also be observed in some of the amateur images. Except for the bright spot 2015-A, which was observed very often, a long-term study of these features requires the use of high-resolution observations acquired by large telescopes to correctly identify most of these features. Even in that case, identification of features visible in amateur images could be wrong if the high-quality observations are acquired with large gaps between consecutive observations.

- Neptune's Southern mid-latitudes had a bright belt of clouds with two major cloud systems in 2013 and 2014. Both cloud systems were approaching one another in 2014 and may have merged before January 2015, when only one bright feature is observed at this latitude range. However, the low number of observations in 2013 and 2014 does not allow to resolve the drift rate of major cloud features in 2013 and 2014 and two sets of drift rates are possible each year. One set of possible motions in 2013 (here called 2013-A and 2013-B) follows the Voyager wind profile. One set of possible motions in 2014 (2014-A* and 2014-B*) also follows the Voyager wind profile and predicts that features 2014-A* and 2014-B* merged in November 2014 shortly before the first detection of spot 2015-A in January 2015.

- The bright spot 2015-A, survived within the same latitude range from January until December 2015 and with the same size and overall shape at least from January until September 2015. HST images at blue wavelengths show a dark feature associated with this bright white cloud. The dark feature SDS-2015 is probably a dark vortex and spot 2015-A is probably a bright white companion cloud linked to the dark vortex. SDS-2015 is similar in some aspects to Neptune's Great Dark Spot (GDS) observed at the time of Voyager 2, but there are also important differences between both systems: The GDS was larger and it was accompanied by a smaller bright cloud system, it oscillated in orientation and shape and drifted in latitude. These are characteristics not observed in 2015 and numerical models will be needed to understand how a massive companion cloud system can accompany a dark vortex in Neptune as in 2015.

- Spot 2015-A and SDS-2015 follow a linear drift close the ambient winds at the latitude of the SDS-2015 (i.e., -45°). The white bright cloud behaves in this sense as a bright companion to a dark vortex which is located slightly further South and drifts with the environment winds. Spot 2015-A has signatures of a possible longitudinal oscillation with a period of 90±3 days and amplitude of 16°±2.5°. These oscillations are consistent with spot 2015-A being a persistent bright companion feature to SDS-2015. Previous dark spots have been seen to oscillate in longitude, so it seems reasonable to assume that the

- oscillations of spot 2015-A track oscillations of SDS-2015. Examples of similar behavior are described by Sromovsky et al. (2001b). Although this is a reasonable interpretation of the data, only one full oscillation was captured and other complex possibilities to explain the changing drift of spot 2015-A could exist.

- The size of the bright spot 2015-A is equivalent to the combined size of major cloud features at nearby latitudes over 2014 (2014-A and 2014-B). However, uncertainties in the drift rates of features over 2013 and 2014 do not allow concluding if spot 2015-A is the remnant of a merger of the large features observed in 2014. The size of spot 2015-A depends on wavelength decreasing in size at shorter wavelengths. This implies that the feature has a compact smaller source at deeper levels that is covered by a high-haze much more extended.

- Bright cloud features at other latitudes, including south polar and north tropical latitudes, were identified on images separated by several days and follow motions similar to the ambient wind as determined from Voyager-2 images. For the polar region, motions compatible with the slow drift of the SPF were found in the ensemble of images. High-resolution HST observations show much faster motions of compact clouds in the elongated polar cloud 2015-P. Those motions also match the Voyager 2 zonal wind profile within large error bars.

- The study of Neptune's atmosphere, particularly the behavior of dark vortices and their effects on the local environment, benefit from strong professional-amateur collaborations. HST observations reveal the location of the dark vortices themselves, ground-based professional observations give precise cloud feature morphology (and infrared photometry for altitude determination), and amateur observations provide a high temporal cadence. Time resolution is crucial for demonstrating feature continuity. For oscillating features like 2015-A, time resolution with high-resolution data from large telescopes, is also essential to prevent aliasing that would arise if oscillation frequencies were determined based on sparsely-sampled observations alone.

**Acknowledgements**

We thank two anonymous referees from their constructive comments that improved the contents of this paper. We are very grateful to Grischa Hahn for his update on the WinJupos software incorporating the ephemeris of Triton that allowed the measurement of amateur images of Neptune. We are also grateful to many amateur observers that observed Neptune intensively over 2013 to 2015 providing data for this research. We are also thankful to P. Irwin for giving permission to use his VLT/SINFONI observations of Neptune. Observations for this research were obtained at the following observatories: Pic du Midi in France, Centro Astronómico Hispano Alemán (CAHA) at Calar Alto, Spain, the Lick Observatory and Palomar Observatory in California and the W.M. Keck Observatory in Hawaii. The Centro Astronómico Hispano Alemán (CAHA) at Calar Alto is operated jointly by the Max Planck Institut für Astronomie and the Instituto de Astrofísica de Andalucía (CSIC). The Robo-AO system was developed by collaborating partner institutions, the California Institute of Technology and the Inter-University Centre for Astronomy and Astrophysics, and with the support of the National Science Foundation under grant Nos. AST-0906060, AST-0960343, AST-0908575, AST-1207891 and AST-1615004, the Mt. Cuba Astronomical Foundation, and by a gift from Samuel Oschin. Research at Lick Observatory is partially supported by a generous gift from Google. The W.M. Keck Observatory is operated as a scientific partnership among the California Institute of Technology, the University of California and the National Aeronautics and Space Administration. The Keck Observatory was made possible by the generous financial support of the W. M. Keck Foundation. The authors wish to recognize and acknowledge the very significant cultural role and reverence that the summit of Mauna Kea has always had within the indigenous Hawaiian community. We are most fortunate to have the opportunity to conduct observations from this mountain. Additional observations were acquired by the Hubble Space Telescope (Programs GO 13937, 14044). Portions of this work were performed under the auspices of the U.S. Department of Energy by Lawrence Livermore National Laboratory under Contract DE-AC52-07NA27344. C.B. acknowledges support from the Alfred P. Sloan Foundation. This work was supported by the Spanish MINECO project AYA2015-65041-P (MINECO/FEDER, UE), Grupos Gobierno Vasco IT-765-13, UPV/EHU UFI11/55 and 'Infraestructura' grants from Gobierno Vasco and UPV/EHU.